\documentclass[lettersize,journal]{IEEEtran}
\usepackage{amsmath,amsfonts}
\usepackage{algorithmic}
\usepackage{array}
\usepackage[caption=false,font=normalsize,labelfont=sf,textfont=sf]{subfig}
\usepackage{textcomp}
\usepackage{stfloats}
\usepackage{url}
\usepackage{verbatim}
\usepackage{graphicx}
\usepackage{cite}
\usepackage[ruled,linesnumbered]{algorithm2e}
\usepackage{cleveref}
\usepackage{array}
\usepackage{xcolor}
\usepackage{longtable}
\usepackage{graphicx}
\usepackage{wrapfig}
\usepackage{units}

\newcommand{\zscomment}[1]{{\color{red} ZS: \sf (#1)}}

\newcommand{\changed}[1]{{\color{orange}\sf #1}}

\begin{document}
\title{Flaky Test Recognition when Testing CPSs Using Hybrid Models}

\author{Zahra Sadri-Moshkenani, Justin Bradley, Gregg Rothermel
\thanks{Zahra Sadri-Moshkenani, Justin Bradley, and Gregg Rothermel are with the Department of Computer Science,
North Carolina State University. E-mail:\{zsadrimo\}@gmail.com and \{zsadrim,gerother\}@ncsu.edu. 
}}

\maketitle

\begin{abstract}
Cyber-Physical Systems (CPSs) have many applications, ranging from simple thermostat systems to autonomous driving systems and medical devices. Like all systems, they need to function correctly. To this end, validation techniques such as testing that can effectively reveal faults are required. Also, CPSs usually operate in uncertain environments where various expected/unexpected events can affect their behaviors. These events together with timing and synchronization incidents may result in various/different CPS behaviors and may cause a CPS to pass a test under some conditions and fail it under others. Such test cases are called ``flaky'' test cases and we call the conditions that are responsible for them ``flaky conditions'' and call these behaviors ``flaky behaviors''. When test cases are flaky, testing results are unreliable. To achieve more reliable test results, engineers may attempt to recognize flaky test cases and remove them. 
In this work, beginning with a test case generation and execution technique called {\sc HyTest} that we created previously, we integrate {\sc TReVa}, a new technique that validates testing results provided by {\sc HyTest} and {\sc FlaRe}, a new technique that recognizes flaky test cases and flaky conditions using hybrid models during the early stages of CPS development in just one additional round of testing. We present the results of an empirical study evaluating the effectiveness of our new approach (which we call {\sc HyTestTF}). Our results show that correctly differentiate flaky test cases from non-flaky test cases

\end{abstract}


\section{Introduction}
\label{sec:intro}

Cyber-Physical Systems (CPSs) integrate software and hardware, communicating through sensors and actuators, typically in a feedback loop~\cite{AlurBook}. They adapt their behavior based on environmental feedback they receive in order to achieve their goals. They are often utilized in safety-critical applications such smart homes, medical devices, and autonomous driving systems; hence, they must meet specific functional and non-functional requirements and operate correctly. Engineers utilize processes such as reviews, testing, and formal methods to ensure that CPSs comply with their requirements~\cite{FisherBook}. This article focuses on testing methodologies, especially test case generation and execution for CPSs' control models and software components. It is important to note that in this context, ``testing'' differs from ``verification''. Verification offers a formal proof of specific aspects of a system's operations across a potentially infinite set of parameters and inputs \cite{KapinskiArticle}, whereas testing assesses whether a system has been developed correctly with respect to its requirements and addresses end-user needs appropriately \cite{Pressman}.\footnote{Sections \ref{sec:Verification} and \ref{sec:FvT} discuss testing and verification further.} 

Where testing of CPSs is concerned, several things require particular attention. First, due to CPSs frequent involvement in safety-critical operations, postponing testing until a CPS is operational in a real-world setting poses risks to both equipment and individuals. Consequently, conducting tests at various stages throughout the development process is imperative. Safety-critical CPSs are typically intricately designed, incorporating a diverse array of model abstractions used during development, which can be examined at different stages. For instance, when evaluating continuous controllers in automotive settings, a tiered testing approach may be employed, including Model in the Loop (MiL), Software in the Loop (SiL), and Hardware in the Loop (HiL) strategies \cite{MatinnejadInpro2}. These levels are explained in section \ref{sec:CPSLevels}. This tiered testing approach ensures frequent testing at critical implementation stages, mitigating the risk of faults slipping through to later deployment phases.

Additionally, CPSs commonly exhibit a blend of continuous and discrete behavior and dynamics. These {\em hybrid} systems are frequently represented as {\em hybrid automata} \cite{HenzingerBook}. Generating test cases that assess both the continuous and discrete behaviors of these systems is essential, regardless of whether engineers are testing simulated or fully implemented CPSs.

Furthermore, Cyber-Physical Systems (CPSs) typically function as {\em reactive systems}, aiming to sustain continuous interaction with the environment amidst various inputs, including time. Consequently, testing must accommodate the potential non-termination of the system \cite{HarelBook}. The standard correctness requirement for such systems stipulates that all executions must be permissible based on system requirements and specifications \cite{CernInBook}. For instance, consider a scenario in which a robot's objective is to reach a goal target via any available path, rather than by following a specific trajectory. In such cases, traditional {\em test oracles}\footnote{A {\em test oracle} is a method or device by which engineers can determine whether or not a test case results in program behavior that conforms to specified behavior (and thus can also indicate whether faulty output or behavior have occurred). For additional details see Section \ref{sec:FvT}.} may prove ineffective at determining trajectory correctness, given the infinite possible trajectories. Instead, the oracle should focus on expected behavior, emphasizing reaching the target while avoiding obstacles, regardless of the specific path. The reactive nature of CPSs has implications for the suitability of test oracles, as traditional ones may be insufficient in this context.

Finally, CPSs usually operate in uncertain and nondeterministic environments where there are many predictable/expected and unpredictable/unexpected events. For example, to an autonomous vehicle, a sudden full stop of a car in front can be expected but likely unpredictable, because it may be interpreted as a slow-down rather than a sudden and full stop. On the other hand, a car driving backwards down a highway is unexpected but possibly predictable, because the backup lights are likely to be on and the distance between the autonomous vehicle and the oncoming car is decreasing. On the other hand, CPSs deal with the timing and synchronizing of processes and devices. Like other systems, a CPS may undergo different process timing and synchronization even over the same test scenario, and these may change the results of executing and testing the CPS. Also, some hardware components may function differently under different conditions, such as changes in external temperature. A CPS that is under test in these various, unpredictable, and changing environments may operate differently and continue to operate correctly or fail when it is tested more than once using the same test case. Such a test case is called a {\em flaky test case}. A ``flaky test case'' is a test case that displays a seemingly random outcome -- pass or fail -- when executed at different times on the same version of the code \cite{BellInPro,EckInpro,LamArticle}. A test case can be flaky when the conditions under which the system is running change unexpectedly, causing the system to act differently. In this work we use the term ``flaky conditions'' to refer to the different conditions that cause the system to act differently and pass or fail on the same test case, and use the term ``flaky behaviors'' to refer to the different behaviors the system may exhibit once it executes under flaky conditions. Flaky test cases can have negative effects on the testing process. Because flaky test cases randomly fail or pass, reports of failures using these test cases can be inaccurate and unreliable and fault localization can be harder. Also, flaky test cases can delay the release time of the final product. 

The most widely-used method for recognizing flaky test cases is ``Rerun'', which involves rerunning each failing test case multiple times to see whether it passes or not. If the test case passes at least once, then the test case is definitely flaky, but if no pass is observed the status of the test case is unknown. There are several disadvantages to using Rerun to identify flaky test cases in a system \cite{BellInPro}. The most important is that flaky test cases are nondeterministic, so there is no guarantee that rerunning a flaky test case will change its outcome and therefore reveal its flakiness. Also, the efficiency of Rerun grows worse with the number of failing test cases, and in the context of CPSs with potentially expensive CPS execution/simulation time, deciding on the number of test case reruns is a challenge. Moreover, re-executing every failed test case is extremely costly when there are large numbers of them or when the system is being tested at the HiL level where any failure may damage the hardware and equipment or endanger human safety. Finally, rerun re-executes test cases that have failed ignores the possibility of test cases that have passed might be flaky -- and they could be.

\textbf{Problem:} To address the first three challenges for testing CPSs that were described above, we have proposed a test case generation and execution technique called {\sc HyTest}\cite{SadriHyTest}. 
{\sc HyTest}\footnote{See Section \ref{sec:HyTest} for more details.} allows {\em pre-MiL level} test case generation, such that the generated test cases can then be used to perform system testing of the simulation models of the CPS at the MiL and SiL levels or of the final product at the HiL level.\footnote{In this work, we focus primarily on the MiL level.} As {\sc HyTest} generates test cases, it also employs an algorithm that reduces the incidence of redundant test cases, based on the hybrid model, rendering testing more efficient. Finally, {\sc HyTest} provides a test oracle that uses the generated test cases to test the CPS as a reactive system. In other words, {\sc HyTest}'s test oracle checks whether the system operates correctly and in an acceptable manner, i.e., over an acceptable order of CPS states and transitions, to meet the goal or not. 

To address the last challenge, in this paper, we target the problem of validating CPS test results, predicting potential flaky test cases, and recognizing real flaky test cases and their corresponding flaky conditions without rerunning the test cases an indeterminate, and theoretically infinite, number of times. 

\textbf{Prior Work:} Several techniques for recognizing flaky test cases in software systems have been proposed \cite{BellInPro,LamInpro}; however, these focus on assessing the possible flakiness of failed test cases and trust passed tests, even though those passes may result from flaky conditions. Also, these approaches have not been applied to CPSs. There are papers, e.g. \cite{ZampettiInpro} that discuss the root causes of flaky behaviors in CPSs and provide some solutions for maintaining the CPSs after flaky test cases are revealed, but they do not discuss techniques for recognizing those test cases. To the best of our knowledge, there is no work on recognizing flaky test cases for CPSs without using Rerun. 

\textbf{Our Technique: } 
As noted earlier, CPSs are usually hybrid systems \cite{HenzingerBook}, and can be mathematically modeled by hybrid models~\cite{TahaBook}. Testing based on hybrid models targets both continuous and discrete CPS behaviors \cite{ANTSAKLISInColl,HenzingerBook}. Further, as also noted earlier, because of the complicated interactions between the cyber 
and physical environments of a CPS and the uncertain environment in which a CPS operates, it is difficult to predict all the incidents (unexpected events) a CPS will encounter during testing \cite{SchneiderInPro}. These incidents affect the CPS's states, which are all captured in the hybrid model as invariants and guard conditions (this is also discussed further in Section \ref{sec:models}.) In other words, the conditions under which a CPS operates at each moment of its execution can be recognized with the help of its hybrid model; therefore, they can be used to recognize the flaky conditions that result in flaky behaviors, and consequently to recognize the potential flaky test cases.  

In this article, we integrate a technique for validating test results, which we call {\sc TReVa}, and a technique for handling flaky test cases, which we call {\sc FlaRe}, into our {\sc HyTest} technique \cite{SadriHyTest}. We call the new technique {\sc HyTestTF}; the technique allows us to validate the test results that {\sc HyTest} provides, to recognize real passes and real failures\footnote{Real passes and real failures are those exhibited by test cases that are not affected by flakiness.}, and to determine flaky test cases and their corresponding flaky conditions in just one additional round of testing. 

While testing the CPS, {\sc HyTestTF} first assesses whether a test case may fail and pass at time $j$ under different conditions and if it determines that it can, marks the test case as a potential flaky test case, and issues a failed test verdict for it. {\sc HyTestTF} records the values that may lead the CPS to pass and fail at time $j$. Next, it re-tests the CPS using the potential flaky test case, while leading the CPS to pass the test,  by setting the values of the CPS's variable at time $j$ to those that cause the CPS to pass the test. If the CPS passes the test, {\sc HyTestTF} recognizes the test case as a real flaky test case; otherwise, it marks the observed failure as a real failure. To validate the test results for those test cases that are not marked potentially flaky, {\sc HyTestTF} retests the CPS using those test cases. This step allows the algorithm to validate whether these failures and passes are the result of flaky conditions in the CPS or not.  After the test results have been validated and any faults discovered have been localized and corrected in the CPS, {\sc HyTestTF} checks whether the potential flaky test cases are real flaky test cases -- i.e., whether they fail under some conditions and pass under different conditions. For each potential flaky test case, {\sc HyTestTF} leads the CPS to pass once, and leads the CPS to fail once. If the CPS passes the test when a pass is expected and fails the test when a failure is expected, {\sc HyTestTF} has revealed a real flaky test case. Otherwise, there is a real failure that needs to be corrected.

Finally, {\sc HyTestTF} recognizes the real flaky test cases that are not revealed in the previous step by recognizing potential flaky conditions with the help of the hybrid model and testing the CPS under those conditions. Just as in the previous step, if the CPS passes the test when a pass is expected and fails the test when a failure is expected, the test case is a real flaky test case, otherwise, there is a real failure. Finally, for those test cases that are recognized as real flaky test cases, {\sc HyTestTF} provides their corresponding flaky conditions. 

This work provides the following contributions:

\begin{itemize}

\item A new set of terminology about, and definitions of flakiness to provide a better understanding of what flakiness is in the context of CPSs.  

\item 
A novel approach for validating test results provided by our previous technique, {\sc HyTest}, for CPSs at the MiL level. 

\item 
A novel approach that recognizes real flaky test cases in just one more round of testing CPSs by checking behaviors that the CPS under test displays, and the states and transitions covered in the hybrid model, at the MiL level.

\item 
A novel approach for recognizing the ``flaky conditions'' under which a CPS may fail or pass a test, with the help of hybrid models at the MiL level.  


\item 
An empirical study examining the application of {\sc HyTestTF} applied to simulation models developed in Simulink \cite{Simulink}
at the MiL level, to measure its effectiveness in recognizing real flaky test cases. Our results show that overall the approach was able to correctly differentiate flaky test cases from non-flaky test cases.  

\end{itemize}

\section{Background}
\label{sec:background}

\subsection{Verification}
\label{sec:Verification}

Verification encompasses tasks aimed at confirming that 
a system accurately executes a designated function; as such,
it confirms that specified requirements have been fulfilled  \cite{Vocab}. 
Verification, usually, formally/mathematically proves that a system,
or some aspect of a system, is correct for a (possibly infinite) 
set of parameters and inputs \cite{KapinskiArticle}. 
Let $M$ be a system, $P$ be an (infinite) set 
of parameters, $U$ an (infinite) 
set of inputs, and $\Psi$ a property that should hold for $M$. 
In such a system, $\Phi(M,p,u)$ is the behavior of $M$ under parameter $p$ 
and input $u$ where $p \in P$, $u \in U$. Also, $\Phi(M,P,U)$  is the set of all 
possible behaviors of $M$ under the parameters in $P$ and inputs in $U$. 
In such a system, verification tries to prove 
that $\Phi(M,P, U) \models\Psi$ for a given set 
of $P$ and $U$ \cite{KapinskiArticle}.

\subsection{ Falsification versus Testing}
\label{sec:FvT}

\textit{{\em Falsification}} techniques search for parameters and 
inputs from (possibly infinite) sets that falsify (violate) 
certain highly-specified properties of systems \cite{HoxhaInPro}.
In system $M$, the goal of falsification is to find a 
$p \in P$ and $u \in U$ such that $\Phi(M,P, U) 
\not\models \Psi$ \cite{KapinskiArticle}.

\textit{{\em Testing}} involves executing a system or component 
under defined conditions, observing or recording the results, 
and subsequently evaluating some aspect of the system or component \cite{Vocab}. 
More precisely, testing is an approach that 
applies specific inputs, events or other stimuli to a system, executes 
the system, and examines system outputs or behaviors for conformance 
to specifications, with the aim of determining whether a property 
holds for a given (finite) set of parameters and inputs \cite{KapinskiArticle}. 
In system $M$, the goal of testing is to find a finite subset 
of values and parameters that hold the property $\Psi$, i.e., 
$\Phi(M,\hat{P}, \hat{U}) \models\Psi$, where $\hat{P}  
\subseteq P$ and $\hat{U}  \subseteq U$ \cite{KapinskiArticle}.  

Falsification involves actively seeking 
parameters and inputs from (potentially infinite) 
sets that contravene the requirements, while 
testing assesses whether a property holds 
for a specific (finite) set of parameters and 
inputs \cite{HoxhaInPro}. 
For system $M$, if falsification cannot find 
an input $p$ that violates a given requirement 
$\Psi$, there is still no guarantee that there 
is a subset of inputs $\hat{P}$ that satisfy requirement $\Psi$. 
Also, if testing determines a subset of inputs 
$\hat{P}$ that satisfy requirement $\Psi$, there 
is no guarantee that there is no input $p$ 
that violates requirement $\Psi$. 
Therefore, testing and falsification are 
complementary.

\subsection{Testing Terminology}
\label{sec:terminology}
Here we briefly present testing terminology that is used in this paper.

Software testing is an approach for assessing whether a the outputs of a software execution, using specific inputs, conforms to specifications and the required properties hold for a 
given (finite) set of parameters and inputs \cite{KapinskiArticle}.

\subsubsection{Test Cases, Test Oracles, and Flaky Test Cases}
\label{sec:testcases}

A {\em test case} is a set of program inputs, execution 
conditions, and expected results developed to assess a particular 
path through the code or a specific behavior of a system~\cite{Vocab}. 
Typically, a test case is a set of test data and preconditions 
that are used to investigate whether the execution of the 
system using the test data and under those preconditions 
leads to expected results and post-conditions.

A {\em test oracle} is a mechanism that compares the result of executing a test case 
with the expected result of running that test case on the system and issues a verdict,
``passed'' or ``failed'', to show the system executed correctly or not~\cite{MemonInPRo}.

A {\em flaky test case} is a test case that displays a seemingly 
random outcome -- pass or fail -- when run at different times on 
the same version of the code \cite{BellInPro,EckInpro,LamArticle}. 
Typically, this occurs because of non-determinism, such as may be 
caused by synchronization or system timing, or due to the uncertain 
situations the system may encounter during the testing process, 
and more generally in the final (virtual or real) environment in 
which the product is supposed to operate.

\subsection{Models}
\label{sec:models}
Models can express the structure and 
behavior of systems through conceptual 
or mathematical representations~\cite{scientificmodeling}. 
Models are useful for describing, developing, and 
validating systems such as CPSs. 
A {\em formal model} is a model that expresses the 
properties of a system at some level of abstraction~\cite{LimerkInPro}. 
Formal models typically represent systems using certain formalisms, e.g., linear temporal logic~\cite{emerson1990temporal}, and are created
prior to the development and deployment of such systems~\cite{NummenmaaArticle}.
A wide range of models have been utilized where software systems are concerned.
Here we describe two types of models that are relevant 
to CPSs and that we use in this work.

\subsubsection{Hybrid Models}

A hybrid automaton is a formal model that 
is used to represent a dynamical system with 
discrete and continuous components \cite{HenzingerBook}. 
A hybrid automaton is a labeled and directed graph 
(a finite state machine) that has the following 
components \cite{ANTSAKLISInColl,HenzingerBook}:

\begin{itemize}

\item  
$X$: A finite set $X = \{x_1,..., x_n\}$ of real-valued variables. 

\item 
$V$: A finite set of vertices, or discrete modes, indicating a control mode/location. 
Traditionally these modes are ``acceptable'' because they are known/expected in the 
CPS and show the possible modes for the CPS while it is functioning. 

\item 
$E$: A set of directed arcs or {\em edges} between vertices. 
An edge  $e \in E$ is also called a {\em control switch} or {\em transition}.

\item 
{\em Flow condition}: equations involving variables in $X$ 
that describe the continuous evolution of the system. 
While the hybrid automaton is in control mode $v\in V$, 
the variables $x_i$ change according to the flow condition.

\item 
{\em Invariant condition}: A condition under which the 
hybrid model may reside in control mode $v$.

\item 
{\em Guard}: An expression involving variables in $X$. 
Each transition is associated with a guard. 
A transition is enabled when its assigned guard is true 
and its execution modifies the values 
of the variables in the hybrid model, as necessary, according to the flow conditions. 

\item 
{\em State}: A state $\sigma = (v, x)$ of the hybrid automaton consists 
of a mode $v \in V$ and a continuous state $x_n\in X$. 

\end{itemize}

In a hybrid automaton model of a CPS, vertices model the 
discrete states, or modes, of the system while edges 
model its discrete dynamics or switches. 
Such an automaton can be used to design a controller 
and to develop a CPS, via simulation models or an implementation. 

Figure \ref{fig:InvertedPendulumHybridModel}, 
taken (with minor modifications) from
\cite{INHybrid}, depicts a hybrid model for a CPS.
(This model and its corresponding CPS  
are described in Section \ref{subsec:CPS}.) 
In Figure~\ref{fig:InvertedPendulumHybridModel}:

\begin{figure}[!t]
\centering
  \includegraphics[width=0.5\textwidth]{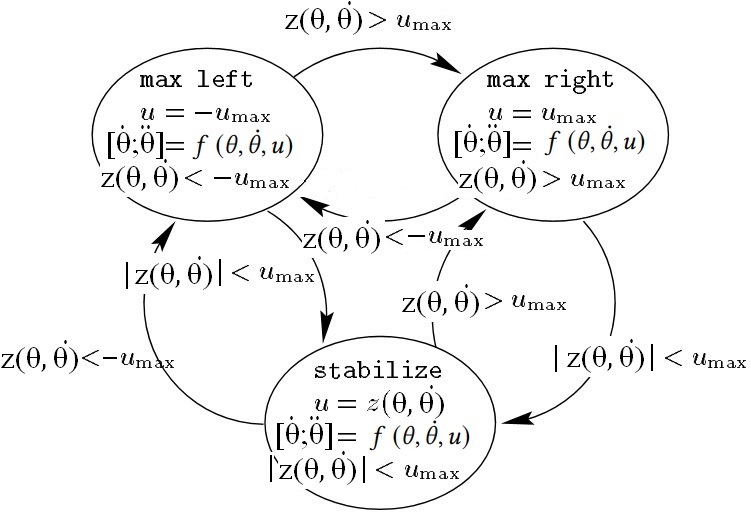}
  \caption{Hybrid model for an inverted pendulum\cite{INHybrid}.}
  \label{fig:InvertedPendulumHybridModel}
\vspace*{-12pt}
\end{figure}

\begin{itemize}
\item
$X=\{\theta, \dot{\theta}, u \}$ 

\item
$V=\{ \textit{``max \ left''}, \textit{``max \ right''}, \textit{``stabilize''}\}$ 

\item
$E=\{ \textit{``max left''} \rightarrow \textit{``max \ right''}, \textit{``max \ right''}\rightarrow \newline \textit{``max \ left''} \textit{``max \ left''}\rightarrow \textit{``stabilize''}, \textit{``max \ right''} \rightarrow \newline \textit{``stabilize''}, \textit{``stabilize''}\rightarrow \textit{``max \ left''}, \textit{``stabilize''}\rightarrow \textit{``max \ right''} \}$ 

\item
Flow conditions: $[\dot{\theta}; \ddot{\theta}]=f(\theta, \dot{\theta},u),  u= z(\theta, \dot{\theta}) \newline \text{ in} \textit{``stabilize''},  
u=-u_{max} \text{ in} \textit{``max left''} ,u=u_{max}\text{ in}  \newline \textit{``max right''}$ (See [12] for more details).

\item
Invariant conditions: $\{ z(\theta, \dot{\theta})<-u_{max}, z(\theta, \dot{\theta})>u_{max}, \newline |z(\theta, \dot{\theta})|<u_{max} \}$ 

\item
Guards: $\{ z(\theta, \dot{\theta})<-u_{max}, z(\theta, \dot{\theta})>u_{max}, |z(\theta, \dot{\theta})|<u_{max} \}$. 

\end{itemize}

\noindent

\subsubsection{Simulation Models}

Simulation models typically form the 
basis of computer simulations~\cite{DuranArticle}. 
Simulation models can be executed in a virtual 
environment to demonstrate the behavior of a
system before that system has been implemented. 
Tests can be performed as simulations
on simulation models~\cite{Simulation}.

\vspace*{-6pt}
\subsection{Model-Based Testing}

{\em Model-based testing} is typically
performed prior to system development and deployment 
allowing test engineers to examine whether a system's model 
conforms to the system's specifications. 
Model-based testing techniques typically generate test cases 
from structural or behavioral models of a system.~\cite{BroyBook}. 
Such {\em model-based test cases} are typically
abstract, and require additional detail relevant
to a system's implementation to be
added to them; this transforms 
them into {\em concrete test cases} that can
be executed on an implemented system 
or its more detailed models. 
Model-based test cases are typically easier to 
maintain than code-based test cases, and 
they can be used to measure the 
{\em coverage} (of the model) achieved in testing.

\subsection{CPS Simulation/Test Levels}
\label{sec:CPSLevels}

CPSs can be typically be simulated and tested
using models at three different levels 
of abstraction~\cite{MatinnejadInpro2}:

\par 1. Model-in-the-Loop (MiL)
\par 2. Software-in-the-Loop (SiL)
\par 3. Hardware-in-the-Loop (HiL)

\noindent
These three levels of abstraction can be described as follows.

\subsubsection{Model-in-the-Loop (MiL)}
 
At the MiL level, CPS simulation is performed in a virtual 
environment, and no physical components are needed. 
Both controllers and plants (physical parts of a CPS) 
are modeled using the same notation and are 
connected in the same diagram.
In many sectors these models are created using Matlab/Simulink~\cite{Matlab,Simulink}, LabView~\cite{LabView}, or other tools that support high-level modeling, and subsequently produce code from the models, e.g., model-based design~\cite{jensen2011model}. 
MiL simulation and testing are performed entirely in a virtual 
environment and without any need for any physical components. 
MiL testing attempts to verify control algorithms, whether 
control algorithms and the plant correctly interact, functional 
requirements, and floating-point arithmetic computations.

\subsubsection{Software-in-the-Loop (SiL)}

Developing and testing at the SiL level is also 
performed in a virtual and simulated environment, 
which often includes the conversion of floating point
data types into fixed-point values as well as the addition 
of hardware-specific libraries~\cite{MatinnejadArticle}. 
Testing at the SiL level is still performed in a virtual 
and simulated environment, as with the MiL level, but 
the focus at the SiL level is on controller code that can 
run on the target platform~\cite{MatinnejadArticle}.
Further, in contrast to verifying behavior, SiL testing 
attempts to verify the accuracy of floating point to fixed-point 
conversions and the conformance of code to control models, 
especially in contexts in which coding may be 
manual~\cite{MatinnejadArticle}.

\subsubsection{Hardware-in-the-Loop (HiL)}

At the HiL level, the controller software is installed on 
the final platform and the plant can be a hardware component,
or software that simulates that hardware component's behavior. 
Testing at this level attempts to verify functional 
and non-functional requirements, the computer executing 
control (e.g., the electronic control unit), 
the real-time infrastructure, and whether the software 
and hardware work correctly together.

\vspace*{-6pt}
\subsection{Extended Example} 
\label{subsec:CPS}

\begin{figure}[!t]
\vspace*{-12pt}
\centering
  \includegraphics[width=0.6\columnwidth]{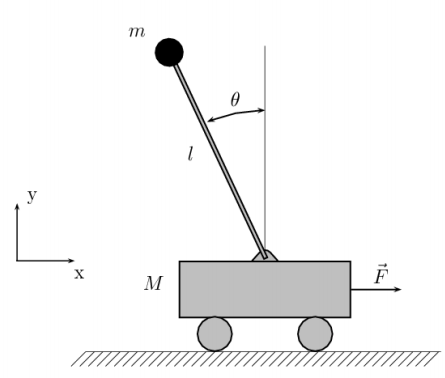}
  \caption{Inverted Pendulum \cite{OlfaInPro}}
  \label{fig:invertedPendulum}
\vspace*{-12pt}
\end{figure}

To illustrate the operations of {\sc HyTest} and 
{\sc HyTestTF} in this article, we utilize the 
simple example presented in \cite{OlfaInPro,INHybrid,SadriArticle}. 
Figure \ref{fig:invertedPendulum} 
depicts a system that includes an inverted pendulum of length $l$ 
and mass $m$ mounted on a cart of mass $M$. 
A force, $F$, is applied to the cart and drags 
it forward or backward to balance the pendulum. 
To maintain stability (i.e., a balanced pendulum), 
a control input is computed and sent to the motors 
in the cart's wheels periodically based on 
feedback the sensors provide, i.e. information about 
the angle, $\theta$, and angular velocity, 
$\dot{\theta}$, of the pendulum.

A general template for a test case for the inverted pendulum
system can be written as follows:
\[
testcase_{i,j}= \left\{ F_i, \theta_i, \dot{\theta}_i, x_i, \dot{x}_i, \theta_j, \dot{\theta}_j, x_j, \dot{x}_j \right\}.
\]

\noindent
In a system such as this, a template 
test case at sampling time, $i$, consists of a 
set of one or more test inputs, (possibly
empty) sets of pre-condition(s), expected 
output(s), and expected post-condition(s). 
In this system, force $F_i$ is the 
test input, while the angle $\theta_i$ and angular 
velocity $\dot{\theta}_i$ of the pendulum, as well 
as the position $x_i$ and velocity $\dot{x}_i$ 
of the cart along the $x$-axis, are pre-conditions. 
The expected test outputs are $\theta_j$ ($j$ can 
be the same as $i$) and $x_j$, and the expected 
post-conditions are $\dot{\theta}_j$ and $\dot{x}_j$. 
Other properties such as the state of the system, 
its failure or success, or safety properties such 
as its stability at a given time $j$, 
could also be expected post-conditions -- we omit
these for simplicity.
A specific test case replaces the variables in the template with concrete values.

The hybrid model for this example, shown in 
Figure \ref{fig:InvertedPendulumHybridModel}, 
has three acceptable modes (shown as ellipses) 
that model its discrete states, and six 
edges/transitions between modes that model 
its discrete dynamics or switches.
Inside each mode, the first line displays the mode's name.  
The next two lines display flow conditions, 
and the fourth displays an invariant condition. 
The labels on each edge display guard conditions. 
The function $z(\theta, \dot{\theta})$, which calculates the total 
energy of the system, together with $u_{max}$, 
which is a constant and shows the maximum total 
energy that the CPS can have in order to remain 
stable, determine whether the CPS should remain 
in the same mode or change to another one. 
In other words, if the absolute value of $z(\theta, \dot{\theta})$ 
is in the range of $-u_{max}$ to $u_{max}$, then 
the inverted pendulum is stable; otherwise it 
is falling down, either left or right.

This hybrid model describes how
the inverted pendulum functions, as follows.
The CPS can begin in any of the acceptable modes. 
Here we assume that the pendulum is initially in the upright 
position, i.e., mode \textit{stabilize}, and when released
the pendulum moves towards the left. 
Suppose force $F$ moves the cart to the right to balance 
the total energy of the CPS and keep the pendulum 
stable, captured by the mode \textit{stabilize}; 
in this case, the pendulum moves to the right. 
Now force $F$ moves the cart to the left to balance 
the total energy of the CPS and stabilize the pendulum. 
If, for some reason, the total energy of the CPS is 
less than $-u_{max}$, i.e., the pendulum is falling 
to the left, the system will be in mode \textit{max left}. 
Depending on the acceptable deviation the controller 
designer has considered for the modes \textit{max left} and 
\textit{max right}, the controller may or may not be able 
to stabilize the pendulum and return the CPS to mode \textit{stabilize}. 

\section{HyTest}
\label{sec:HyTest}

In this section, we summarize our previous work on {\sc HyTest}, which is a test case 
generation technique that generates test cases based on hybrid models, accompanied by
appropriate test oracles, for use in testing CPSs early in their development cycle.
We limit ourselves to text that we believe is necessary to understand our new technique.
Complete details on the {\sc Hytest} algorithms and examples of their operation
can be found in \cite{SadriHyTest}

\subsection{The CPS Development Process} 
\label{sec:CPSDevProcess}

Before introducing {\sc HyTest}, it's essential to understand the CPS development process in which it is intended to be applied. Irrespective of the chosen development methodology, the CPS development process typically begins with a ``Requirements Engineering'' step. Here, the system's goals, properties, specifications, and requirements are established. For CPSs, this step often involves determining and modeling the system's dynamics, usually through mathematical models like equations of motion. Once these models are defined, the hybrid model of the CPS can be designed without additional effort beyond what's already required in constructing the mathematical models and the requirements engineering process.

The subsequent step in the CPS development process involves designing and testing the controller, leading to the integration of the controller's dynamics into the system. At this juncture, {\sc HyTest} utilizes the hybrid model, the system's dynamics, and other information extracted during the requirements engineering process to generate test cases. These test cases are then used to test the simulation model at the MiL level, providing results that identify any failures. Engineers correct these faults, and then {\sc HyTest} is reapplied to ensure the model passes all test cases. This iterative testing and correction process continues until no test case fails. At the SiL level, the controller's code is developed and unit tested, with {\sc HyTest} employed to test the CPS using the same generated test cases. Testing and adjustment continue until no test case fails, ensuring the CPS's readiness for deployment in the real environment, where {\sc HyTest} can also be utilized for testing. Importantly, {\sc HyTest} can generate test cases regardless of the CPS development methodology, offering flexibility and consistency in testing approaches.

Checking the correctness of the hybrid model, including reachability and other issues \cite{HenzingerArticle}, before using {\sc HyTest}, is not addressed in this work. However, in observed cases where the hybrid model was flawed, {\sc HyTest} detected errors, displayed error messages, and terminated execution. Additionally, as {\sc HyTest} assumes every state can be the CPS's start state, reachability checking of the hybrid model to ensure each state is reachable from a specific start state is unnecessary.

\subsection{Overview of {\sc HyTest}} 
\label{sec:HyTestOverview}

\begin{figure*}[t]
\centering
  \includegraphics[width=5in]{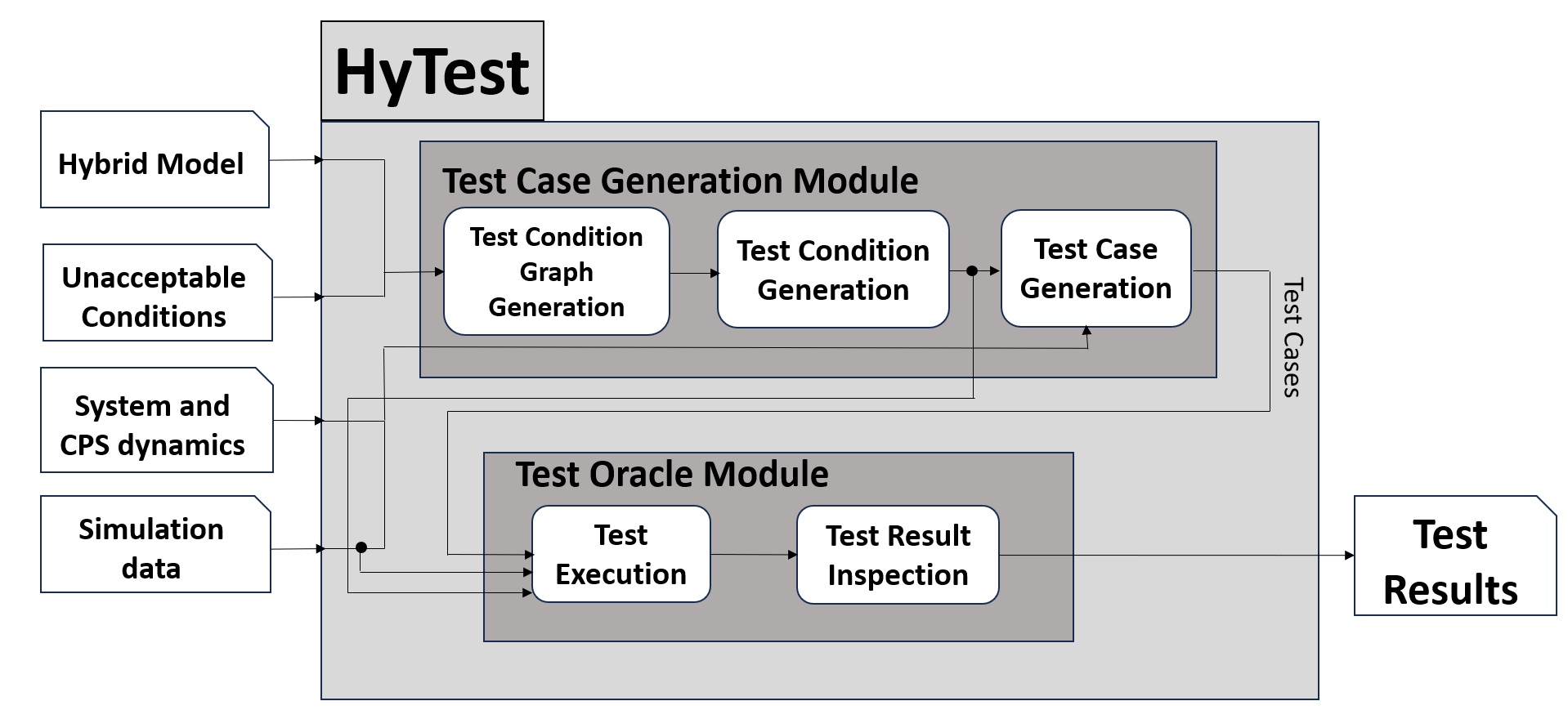}
  \caption{{\sc HyTest} Overview}
  \label{fig:hytestoverview}
\vspace*{-12pt}
\end{figure*}

Figure \ref{fig:hytestoverview} 
provides an overview of {\sc HyTest}. 
As depicted in the figure, {\sc HyTest} consists of two 
main modules: test case generation and test oracle modules. 

The Test Case Generation Module receives data about the 
hybrid model, unacceptable conditions, system and 
CPS dynamics, and simulation data---these data are 
usually available after the requirement engineering phase 
or can be derived easily from the outcome of that phase---in 
the form of inputs that the user provides to answer 
the questions that the module asks and in the form of text files. 

Using this data, the Test Condition Graph Generation step
generates a ``Test Condition Graph'' that shows all 
possible modes and transitions (whether acceptable or not), 
in the form of passing (if they are acceptable) 
or failing (if they are unacceptable) conditions. 
Because the graph includes all acceptable or unacceptable 
conditions, it captures the effect of any event (whether 
expected or not) that may occur in the CPS, 
whether or not the event is known a priori. 

Next, the Test Condition Generation step combines
each condition in the test condition graph with other 
information about the conditions, such as the mode of 
the CPS for each condition, creating a set of test conditions. 
This step generates test conditions for all 
conditions of the transitions and the modes, in order to
achieve complete coverage of the acceptable and unacceptable conditions.

When test conditions are ready, the Test Case
Generation step uses these test conditions to 
partition the input space, i.e., it classifies 
the inputs based on test conditions, and generates test cases. 

After test cases have been generated, the Test Oracle 
Module receives them along with the information about 
the CPS's variables and simulation data, such as simulation time. 
Using these, the module performs a Test Execution step on the CPS, 
followed by a Test Result Inspection step, which checks the 
execution outputs in order to issue the test verdicts. 

Algorithm \ref{alg:approachOverview} 
provides an overview of {\sc HyTest}. 
{\sc HyTest} receives information about a CPS's hybrid 
model, conditions that indicate the CPS's failure, 
simulation parameters, information on the CPS's dynamics, 
and a simulation model of the CPS.
All of these are 
either required to design and develop the 
CPS or can be obtained easily from its requirements. 
The algorithm outputs the failures that are 
revealed during the CPS testing process.

\SetKwComment{Comment}{/* }{ */}
\SetKwInput{KwData}{Inputs}
\SetKwInput{KwResult}{Output}

\begin{algorithm}[h]
\small
\caption{{\sc HyTest}}\label{alg:approachOverview}
\KwData{hModel: Hybrid Model, \newline
unConditions: Unacceptable Conditions, \newline
simData: Simulation Data, \newline
sysDyn: System and CPS Dynamics, \newline
simulationModel: Simulation model}
\KwResult{failures: Detected failure in the CPS}

hModel=getHybridModelData();\label{alg:11}

unConditions= getUnacceptableConditions();\label{alg:12}

simData= getSimulationData();\label{alg:13}

sysDyn= getDynamics();\label{alg:14}

cGraph= genConditionGraph(hModel,unConditions);\label{alg:15}

tConditions= genTestConditions(cGraph);\label{alg:16}

testcases=genTestCases(tConditions,simData,sysDyn);\label{alg:17}

[failures]=testCPS(testcases,simData,tConditions,simulationModels);\label{alg:18}

\KwRet{$failures$}

\end{algorithm}

{\sc HyTest} begins (line~1) by retrieving 
data about the hybrid model of the CPS.
This data includes the total number of modes, 
the CPS goal(s), invariant conditions, guard 
conditions that show the transitions between modes, 
and variables in the hybrid model including 
their acceptable value range and precision. 
Next (line~\ref{alg:12}), {\sc HyTest} obtains a list 
of ``unacceptable conditions'', provided by the 
system designer or test engineer: these are conditions 
that are not expected to occur during the operation 
of the CPS, or that would lead to a failure. 
If no unacceptable conditions are provided for the CPS, 
{\sc HyTest} temporarily sets the variable {\em unConditions} to null
until a later set of operations can adjust it. 
{\sc HyTest} next (lines~\ref{alg:13}-\ref{alg:14}) retrieves 
data about the simulation model used to test 
the CPS, and about the system's and CPS's dynamics
including its state-space representation; transfer 
functions or Matlab code that implements the 
system's dynamics; and initial values, inputs, 
and simulation time, provided as Matlab ``init'' files.
Using this information, {\sc HyTest} creates 
a ``condition graph'' (line~\ref{alg:15})
and using this graph, generates 
test conditions (line~\ref{alg:16}). 

{\sc HyTest} uses test conditions to partition the 
input space and then generate test cases (line~\ref{alg:17}). 
It passes these test cases, along with simulation data,
to the testCPS function (line~\ref{alg:18}).
The testCPS function implements a testing framework 
with which to test the CPS by running its simulation 
model, and ultimately returns the failures that are 
revealed by test cases during testing.  

\begin{figure*}[t]
\centering
  \includegraphics[width=\columnwidth]{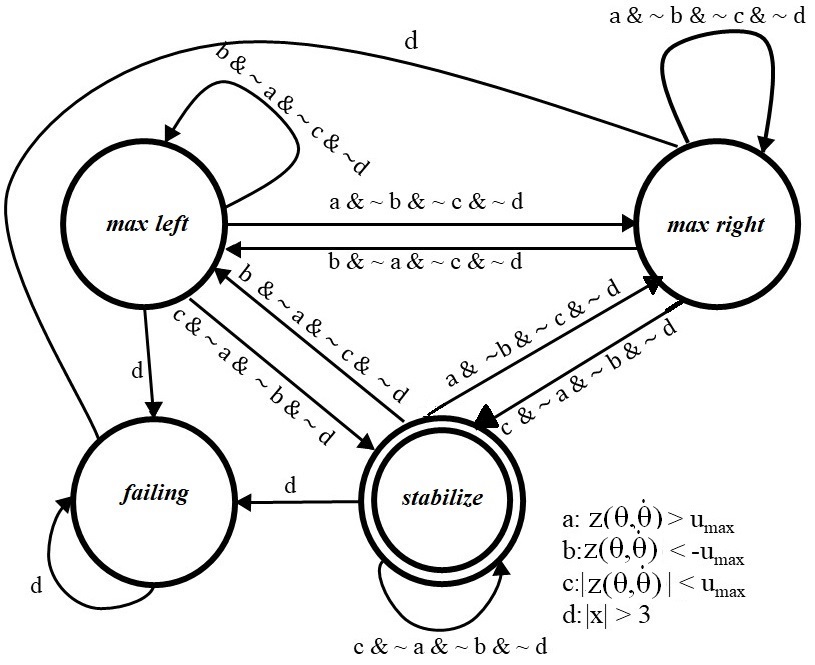}
  \caption{Condition Graph Generation, Step 2}
  \label{fig:substep2}
\vspace*{-6pt}
\end{figure*}

To generate test cases using information about 
the hybrid model and unacceptable conditions, 
{\sc HyTest} generates a specific type of 
graph that we call a {\em condition graph}. 
A condition graph is a representation of a CPS 
that shows all acceptable, final, and failing modes 
in the CPS, along with all conditions that change 
the mode of the CPS from one to another and 
may lead the CPS to a failure or success. 
An example of condition graph is shown in Figure \ref{fig:substep2}. 
As we see, {\sc HyTest} categorizes the 
modes as acceptable (any mode that is neither 
final nor failing), final, or failing, to show whether 
the CPS has reached its goal (final modes), has failed 
(failing mode), or is working toward its goal (acceptable modes). 
As an example, the edge from the mode ``stabilize'' to ``stabilize'' 
shows that if the absolute/value of the total energy of the CPS, 
$z(\theta, \dot{\theta})$, is less than the  maximum total 
energy that the CPS can have in order to remain 
stable, $u_{max}$, and the total energy of the CPS is neither 
greater that  $u_{max}$ nor less that  $-u_{max}$ and 
the displacement of the cart is not more than $3 m$, 
which here is an unacceptable condition, then the CPS's 
stays in the mode ``stabilize''.

For each edge in the graph, {\sc HyTest} obtains its source and destination modes 
and concatenates them as a formatted string, which is easy to parse
when the approach needs to extract specific 
data from the test conditions. 
{\sc HyTest} concatenates the edge's label, which 
shows under what conditions the CPS changes 
mode from source to destination, to the formatted string. 
Finally, it appends the type of the destination mode (of the edge) 
to the formatted string, which helps the test oracle issue the 
correct test verdict for the test inputs. 
If the destination mode is: 1) final then the type is ``passed'', 
2) failing then type is ``failed'', and 3) acceptable then the type is ``acceptable''. 
The following line illustrates a sample test condition that is generated by 
{\sc HyTest} for our example and corresponds to the self-loop on the mode \textit{stabilize}.

\vspace*{6pt}
\noindent
$\textit{stabilize},\textit{stabilize} \ \# \ (|z(\theta, \dot{\theta})| < u_{max}) \ \& \sim(z(\theta, \dot{\theta}) > u_{max}) \ \& \sim(z(\theta, \dot{\theta}) < -u_{max} ) \ \& \sim(x > 3)@passed$

\vspace*{6pt}
Using these test conditions, {\sc HyTest} partitions the input space. 
It examines the test inputs against the test conditions and finds 
all test conditions a test input may fit. 
Each test input may lead the CPS to each of failing, final, 
or acceptable modes, which is recognizable using the type 
of the destination mode in the test condition they fit.

A concrete test case for our inverted pendulum example looks like this:

$testcase_i=$ 
$\{-0.495502602162482, 2.88465092291609, $ \newline \indent
$ 0.529831722628956, 1.33637906750083, passed \}$.

\noindent
The members of this set are the pendulum's angle, 
the pendulum's angular velocity, the cart position, 
the cart velocity, and the expected type of the mode 
that the CPS will be in when the its variables are 
set to these values (in this case the type of mode 
is ``passed'' which shows the CPS starts from a final mode). 

Using these test cases, {\sc HyTest} begins the test execution process. 
{\sc HyTest} starts the CPS in different states, as starting 
state of the execution, and checks whether the CPS operates correctly or not. 
To this end, first {\sc HyTest} sets the value of the CPS's variables 
to the values in a test case and starts the CPS simulation. 
As an example, using the concrete test case for our inverted 
pendulum CPS, {\sc HyTest} starts simulating the CPS from a 
``passing'', i.e., final, state by setting the initial values of the 
CPS's variables to the values in the test case and 
simulates the CPS for the entire simulation time. 
Then, it gets the test output values and checks what 
test condition they fit and recognizes the type of the 
states of each moments of the CPS's simulation. 
Using this information {\sc HyTest} issues the test verdict.

\section{Terminology}
\label{sec:term}

\begin{figure*}[b]
\centering
  \includegraphics[width=5in]{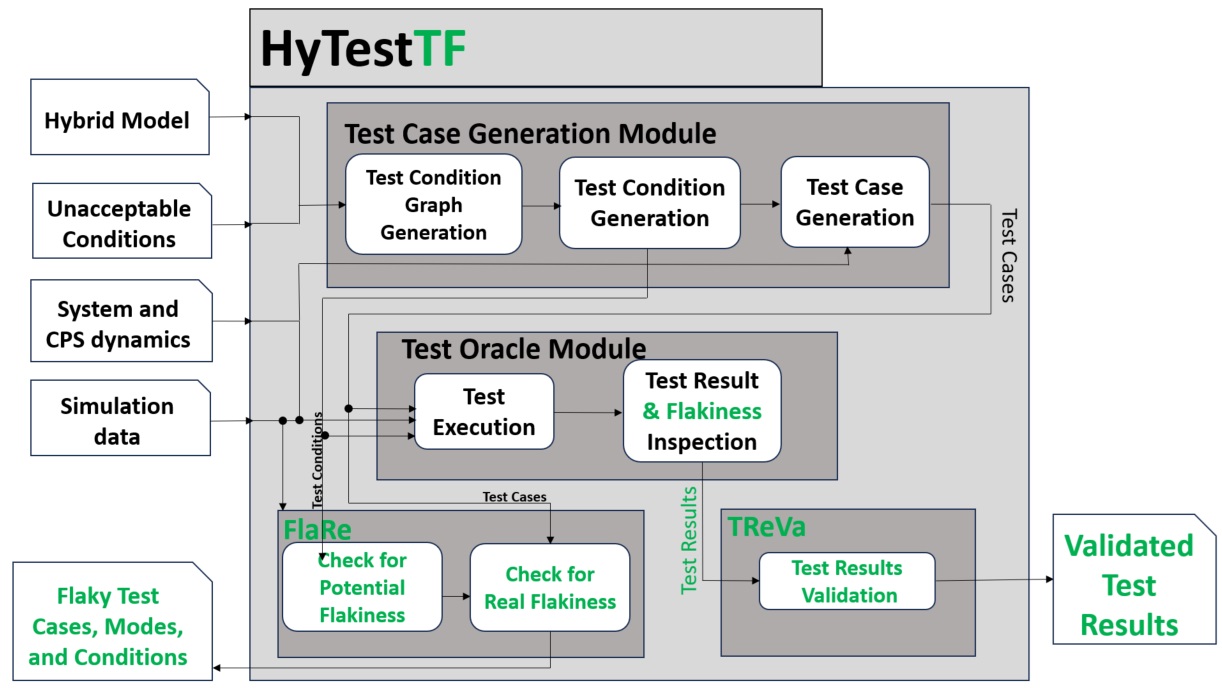}
  \caption{{\sc HyTestTF} Overview}
  \label{fig:overview}
\vspace*{-12pt}
\end{figure*}

We next introduce and define some new terminology 
which we use in explaining our approach. 

\subsection{Flaky Condition}
Flaky test cases are those test cases on which the CPS can randomly fail or pass. 
This happens because of a change in the condition under which the CPS is functioning at some sampling time $j$. We call this condition a {\em flaky condition}. If the condition causes the CPS to fail the test we call it a {\em failing flaky condition}. Otherwise, it is a {\em passing flaky condition}. For example, in the test condition graph that {\sc HyTest} has generated for the inverted pendulum (see Figure \ref{fig:substep2}), for mode ``stabilize'', $ c \ \& \sim a \ \& \sim\ b \ \& \sim d$ which is 
$(|z(\theta, \dot{\theta})| < u_{max}) \ \& \sim(z(\theta, \dot{\theta}) > u_{max}) \ \& \sim(z(\theta, \dot{\theta}) < -u_{max} ) \ \& \sim(x > 3)$ is a {\em potential} passing flaky condition, and $d$, which is $sim(x > 3)$, is a {\em potential} failing flaky condition. We use term {\em potential} here because according to the test condition graph, the CPS may fail or pass when it is in the mode ``stabilize''. Further steps are required to recognize it as a {\em real} passing/failing flaky test condition, as we explain in Section \ref{sec:approach}. 

\subsection{Flaky Behavior}

As noted earlier, CPSs are reactive systems, i.e., they function according to the feedback they receive through their sensors and from the environment. When an incident (an unexpected event) happens, whether within the CPS or in its functioning environment, the CPS reacts to it and adapts its behavior to the condition related to that incident. If the condition is a flaky one, no matter whether passing or failing, we call the CPS adaptation a {\em flaky behavior}. In other words, a flaky behavior is the CPS's reaction to a flaky condition. In Figure \ref{fig:substep2}, when the CPS is in mode ``stabilize'', if a failing flaky condition is encountered the CPS may (we use the word ``may'' because these flaky conditions are just potential flaky conditions) change the mode to ``failing'' and if a passing flaky condition occurs the CPS may stay in the mode ``stabilize'', which is passing. These behaviors are flaky behaviors because they are the results of flaky conditions. 

\subsection{Flaky Modes}

As noted before, hybrid models and the test condition graph that {\sc HyTest} generates have several modes. If the CPS displays flaky behavior while in any mode, we call that mode a {\em flaky mode}. In Figure \ref{fig:substep2} the states ``stablize'', ``max left'', and ``max right'' are potential flaky modes. 

\section{Approach}
\label{sec:approach}

In this section we explain our Test Results Validation ({\sc TReVa}) and Flaky Test Recognition ({\sc FlaRe}) approaches, which as noted before, are integrated into {\sc HyTest} to form {\sc HyTestTF}. 

\subsection{Overview of Approach} 
\label{sec:overview}


Figure \ref{fig:overview} provides an overview of {\sc HyTestTF} and how {\sc TReVa} and {\sc FlaRe} are integrated into {\sc HyTest}. As shown, {\sc TReVa} depends on the module {\em Flakiness Inspection}, which is integrated into {\sc HyTest}'s test oracle. While {\sc HyTestTF} is testing the CPS, it marks those test cases that fail but, under different conditions, have a chance to pass, as potential flaky tests. Next, it sends test results, which include results from potential flaky test cases as well, to {\sc TReVa}. {\sc TReVa} validates the failures and passes by distinguishing real failures and passes from flaky ones. {\sc FlaRe} receives test conditions and test cases. First, using test conditions, it finds all modes that could be flaky. Next, using test cases, it examines the potential flaky modes and checks whether any of them can be real flaky modes. For each mode it checks whether it passes and fails the test case under passing and failing flaky conditions. If so, the test cases that are related to those flaky conditions are real flaky test cases, the mode is a real flaky mode, and the flaky conditions are real flaky conditions under which the CPS may fail while it is operating in the real environment. Using this information, the CPS developer can take preventative steps to avoid failures that occur due to failing flaky conditions.

\vspace*{-6pt}
\subsection{{\sc HyTestTF}'s Test Oracle and Flakiness Inspection}
\label{sec:oracle}

The final step of {\sc HyTest}'s Algorithm~\ref{alg:approachOverview} (line~\ref{alg:18}), presented in Section \ref{sec:HyTestOverview}, invokes algorithm testCPS, into which the flakiness inspection module is integrated. testCPS executes generated test cases on the CPS and inspects the test results and flakiness.

Algorithm \ref{alg:testFramework} presents a modified version of testCPS, with the original lines shown in black, and the new, added flakiness inspection steps shown in green.

\SetKwComment{Comment}{/* }{ */}

\begin{algorithm}[b!]
\footnotesize
\caption{Test execution and test oracle}\label{alg:testFramework}

  \SetKwInOut{Input}{inputs}
  \SetKwInOut{Output}{output}
  \SetKwProg{testCPS}{testCPS\'}{}{}

  \testCPS{$(testcases,simData,tConditions,simModel)$}{
    \Input{testcases: Test Cases,
    \newline simData: Simulation Data,
    \newline tConditions: Test Conditions,
    \newline simModel: Simulation Model of CPS}
    
    \SetKwProg{try}{try}{:}{}
    \SetKwProg{catch}{catch}{:}{end}
    \Output{results: Test results}

    testInputs=extractTestInputs($testcases$); \label{alg:51}
    
    initialMode= extractInitialMode($testcases$); \label{alg:52}
    
    \ForEach{test input $input_i \in testInputs$}{\label{alg:530}

        \color{green!65!black}
            $results.values \gets input_j$ \label{alg:531}
            
            $results.simTime \gets simData.simTime$ \label{alg:532}
            
            $results.modes \gets initialMode_i$ \label{alg:533}
        \color{black}
    
    \try{}{
    
      testOutput=simulateModel($input_i,simData$); \label{alg:53}
      
      possibleCPSModes=recognizeCPSModes(...
      ...$testOutput,tConditions$); \label{alg:54}
      
      \ForEach{mode $mode_j \in possibleCPSModes$}{\label{alg:550}
      
          \uIf{$mode_j == null$}{\label{alg:55}
          
              print(``The hybrid model was not designed correctly.'');\label{alg:551}
          }
          \uElse{
            \uIf{\texttildelow isAllowed($mode_j$, $mode_{j+1}$)}{\label{alg:56}

            \color{green!65!black}{
                $results_i.case \gets ``failed''$ \label{alg:561}
                }
                
            }
            \uElseIf{$mode_j$ contains ``failing''}{\label{alg:57}
            
            \color{green!65!black}{
            \uIf{$mode_j$ \texttildelow contains ``passing''}{\label{alg:570}
            \color{green!65!black}{
                 $results_i.case \gets ``failed''$ \label{alg:571}
            }
             }
            \uElseIf{$mode_j$ contains ``passing''}{\label{alg:58}

                $results_i.case \gets ``flaky''$ \label{alg:580}
                                
                $results_i.values \gets values_j$ \label{alg:581}
                

                $results_i.simTime \gets simData.simTime - j$ \label{alg:582}

                $results_i.modes \gets mode_j$ \label{alg:583}
                }
            }
            }
            
         }
        }
        \uIf{$possibleCPSModes$ \texttildelow contains ``failing''}{\label{alg:59}
        
            $lastMode \gets the last mode\ in \ possibleCPSModes$;
        
            \uIf{ $lastMode$ contains final}{
            
                \uIf{$initialMode_i ==  final \ ||\  initialMode_i == acceptable$}{
                \color{green!65!black}{
                
                    $results_i.case \gets ``passed''$ \label{alg:590}
                }

                }
                \uElse{
                    \color{green!65!black}{
                    
                        $results_i.case \gets ``flaky''$ \label{alg:60}
                        
                        $results_i.values \gets input_i$ \label{alg:601}
                    
    
                        $results_i.simTime \gets simData.simTime$ \label{alg:603}
    
                        $results_i.modes \gets initialMode$ \label{alg:604}
                    }
                    
                }
            }
            \uElse{
                
                \color{green!65!black}{
                $results_i.case \gets ``flaky''$ \label{alg:62}
                
                $results_i.values \gets values_{endTime}$ \label{alg:621}
                

                $results_i.simTime \gets simData.simTime$ \label{alg:623}

                $results_i.modes \gets mode_{endTime}$ \label{alg:624}
                }
                
            }
       }   
      }\catch{Exception}{\label{alg:63}
        \color{green!65!black}{
            $results_i.case \gets ``failed''$\label{alg:631}
        }
      }
      
    }
    \KwRet{$results$}\;
  }

\end{algorithm}

The new testCPS algorithm, referred to here as testCPS\' \ , takes the generated test suite, test conditions, and simulation data as inputs and returns test results, including passes, failures, and potential flaky test cases revealed during testing. testCPS\' \ begins by retrieving test cases and 
extracting test inputs (line~\ref{alg:51}) 
and the initial modes (line~\ref{alg:52}). 

{\sc HyTestTF} starts the CPS in different modes 
and transitions. To this end, for each test case, 
testCPS\' \ (lines \ref{alg:530}-\ref{alg:53}) initially 
sets the values of the variables to the test 
input values of test cases to put the CPS in 
every possible acceptable, final, or failing 
mode and then it monitors the behavior of 
the CPS until the end of the simulation. As an example, using the concrete test case for our 
inverted pendulum CPS (presented in Section \ref{sec:HyTestOverview}), 
{\sc HyTestTF} starts simulating the CPS from a ``passing'', 
i.e., final, mode by setting the initial values of the 
CPS's variables to the values in the test case, and 
simulating the CPS for the entire simulation time. Also, for each test input on lines \ref{alg:531}-\ref{alg:533}, it initializes $results$ of each test run that will be used by {\sc TReVa} to validate the test results.

For each test input, testCPS\' \ receives the simulation 
output(s) as one or more sets of values with the 
length of the (simulation time)/(sampling time). 
Then, testCPS\' \ inspects the simulation outputs 
to see whether the CPS has stopped at a final mode 
through sequences of allowed transitions and whether
the values of the variables fit the goal conditions. 

The test oracle behaves as follows. 
First, using the test conditions and the test output 
signal, the test oracle recognizes the possible modes 
of the CPS at any sampling time (line \ref{alg:54})
by checking the test outputs against test conditions. 
Since a test output at time $j$ may fit in several test 
conditions, the CPS at time $j$ can be in different modes 
and the real mode of the CPS depends on its real condition at testing time. 
At this point (lines \ref{alg:550}-\ref{alg:551}), testCPS\' \ checks whether 
there are any test outputs with no modes assigned to them. 
If there are, the hybrid model was designed incorrectly (line \ref{alg:551}),
because there is a behavior in the CPS that does not 
fit any modes/transitions represented in the hybrid 
model: i.e., the hybrid model omitted a mode/transition. 
According to the modes in the condition graph, the 
modes and transitions (based on their destinations) 
of the CPS can be categorized into three 
groups: acceptable, final, and failing. 

When the possible modes the CPS can be in at a sampling time are recognized, the flakiness inspection part of the test oracle inspects the test results to find test cases that may be flaky. To this end, it looks for one of the following cases:

\begin{itemize}
    \item At any sampling time $j$, the CPS has more than one possible mode; i.e., the test outputs at sampling time $j$ fit in more than one test condition, and the set of possible CPS modes contains at least one ``failing'' and at least one ``passing'' mode (lines \ref{alg:58}-\ref{alg:583}).

    \item The CPS has passed the test while we expect a failure (lines \ref{alg:60}-\ref{alg:604}). In this case, if the CPS is retested for the whole simulation time while it starts under a flaky failing condition, the test may fail instead of pass, which is expected. 

    \item The CPS has not been in any ``failing'' mode during the whole simulation time, and we do not expect the CPS to fail the test, but the CPS was not able to reach its target, i.e. the CPS's mode at the end of its simulation is ``acceptable'' and not ``final''. In this case, although the CPS has failed the test, it may pass it if it is given more time (lines \ref{alg:62}-\ref{alg:624}). 
    
\end{itemize}

If any of these cases occur, the test oracle marks the test case as a ``potential flaky test case'' and saves the values at sampling time $j$, the rest of simulation time, and the possible modes of the CPS at time $j$. Using this information, {\sc TReVa} can start the CPS from the state in which the test oracle has recognized the flakiness and continue testing the CPS for the rest of the simulation time to see whether the CPS passes or fails the test and whether the failure is a real or a flaky one.   

Based on the potential cases of flakiness, to issue the correct test verdict for each test case, the test oracle works as follows:

\begin{itemize}

\item 
Using the condition graph, the test oracle determines 
(lines \ref{alg:56}-\ref{alg:561})
whether transitions between consecutive modes are 
allowed or not, i.e., transitions between 
modes of the test output at times $i$ and $i+1$. 
If a transition is not allowed then a failure 
occurred, e.g., there is an overshoot, (a sudden change in a value at two consecutive sampling times) in the CPS output values.

\item 
If the possible modes of the CPS include ``failing'' 
in at least one sampling time and do not contain ``passing'' at the same sampling time, the test oracle issues 
a ``failed'' verdict for the test case (line \ref{alg:57}-\ref{alg:571}), but if they contain ``passing'' at the same sampling time, then the failure, or the pass, or in general the test run, is ``flaky'' (lines \ref{alg:58}-\ref{alg:583}).
    
\item 
If the test verdict is not ``failed'', the possible modes 
of the CPS at the last moment of the simulation include ``final'',  
and the initial mode in the test case includes ``final'' or ``acceptable'', 
then the test oracle issues a ``passed'' verdict (lines~\ref{alg:59}-\ref{alg:590}). 
    
\item 
If the test verdict is not ``failed'', the possible modes of 
the CPS at the last moment of the simulation includes ``final'', 
and the initial mode in the test case includes ``failing'', 
then the test oracle issues a ``flaky'' verdict (lines~\ref{alg:60}-\ref{alg:604}).
    
\item 
If the test verdict is not ``failed'' and the possible modes 
of the CPS at the last moment of the simulation include ``acceptable'', 
then the CPS has failed the test (lines \ref{alg:62}-\ref{alg:624}), 
because the CPS was not able to reach its target, i.e. the 
final mode or goal conditions, within the simulation time 
(the time within which the CPS must reach the goal condition), but since the CPS may reach the goal in a longer time and pass the test, so the test oracle issues a ``flaky'' verdict. 
In our inverted pendulum system, although the modes 
``max left'' and ``max right'' are both acceptable, 
the goal of designing such a CPS is to keep it in the 
``stabilized'' mode; therefore, if this CPS is not in 
the mode ``stabilized'' at the end of the simulation, 
the CPS failed to reach its goal. 
In this case, although the CPS is in an acceptable 
mode at the end of the simulation, {\sc HyTestTF} recognizes 
a failure because the goal was not reached. 

\item
If the simulation encounters any problem 
that leads to an exception, the test oracle 
issues a ``failed'' verdict (lines \ref{alg:63}-\ref{alg:631}).

\end{itemize}

\vspace*{-6pt}
\subsection{{\sc TReVa}: Test Results Validation}
\label{sec:treva}

After testCPS finishes the testing and returns the test results, {\sc TReVa} (Algorithm~\ref{alg:treva}) receives the results, test conditions, and simulation data and model as inputs and returns validated test results, which include real passes, real failures, and real flaky test executions. 

\SetKwComment{Comment}{/* }{ */}

\begin{algorithm}[t]
\small
\caption{TReVa}\label{alg:treva}

  \SetKwInOut{Input}{inputs}
  \SetKwInOut{Output}{output}
  \SetKwProg{TReVa}{TReVa}{}{}

  \TReVa{$(results,simData,tConditions,simModel)$}{
    \Input{results: Test Results,
    \newline simData: Simulation Data,
    \newline tConditions: Test Conditions,
    \newline simModel: Simulation Model of CPS}
    
    \SetKwProg{try}{try}{:}{}
    \SetKwProg{catch}{catch}{:}{end}
    \Output{validatedResults: Validated Results}

    [testcases, simData]=extractVlaues($results,...$
    
    $...newSimData$); \label{alg:70}

    [newResults, $\sim$ ]=testCPS($testcases,newSimData,...$\label{alg:71}
    
    $tConditions,simModel$);\label{alg:72}

    \ForEach{new result  $newRes_i \in newResults$}{ \label{alg:73}

        \uIf{$newRes_i.case$ == ``passed'' \&\& $results_i.case$ == ``passed''}{\label{alg:74}
            $validatedResults_i \gets$ ``passed'' \label{alg:75}
        }
        \uElseIf{$newRes_i.case$ \texttildelow= ``passed'' \&\& $results_i.case$ \texttildelow= ``passed''}{\label{alg:76}
            $validatedResults_i \gets$ ``failed'' \label{alg:77}
        }\uElse{
            $validatedResults_i \gets$ ``flaky'' \label{alg:78}
        }
    }
    
    \KwRet{$validatedResults_i$}\;
  }

\end{algorithm}

{\sc TReVa} begins by retrieving values and start modes from the set of results (line \ref{alg:70}). Using the values and start mode, it generates a new set of test cases. {\sc TReVa} uses this new test suite to start testing the CPS from the states in which flakiness has been recognized for the test cases in the original test suite, which was generated by {\sc HyTest}, for those test runs that are marked  potentially flaky. Also, for each potential flaky test case, {\sc TReVa} updates the simulation time to the value that is retrieved from it. for the test runs that are not marked potentially flaky, {\sc TReVa} uses this new test suite to validate the results for those test runs. 

Once {\sc TReVa} has the new test suite and updated simulation time for each test case, it calls testCPS (line \ref{alg:71}-\ref{alg:72}) to retest the CPS using the new test suite. When testCPS returns the new results, in lines \ref{alg:73}-\ref{alg:78}, {\sc TReVa} compares them with those results it received from the test oracle. If the CPS has passed the test both times (lines \ref{alg:74}-\ref{alg:75}), then {\sc TReva} issues a passed test verdict for the test case. If the CPS has not passed the test both times, i.e. it has failed the test or the test run is a potential flaky one, (lines \ref{alg:76}-\ref{alg:77}), then {\sc TReva} recognizes the test case as a real failed one. Remember that a potentially flaky test run is a test run that has failed, but it has a chance to pass. On line \ref{alg:78}, if the CPS has not passed the test either times and passed it the other time, then {\sc TReVa} marks the test case as a real flaky test case. 

It is worth mentioning that the faults that caused the real failures may be the cause of flaky behaviours in the CPS, too. and fixing them may resolve the flaky behaviours, as well; Therefore, at this stage, a test engineer should resolve the real failures and ignore those that are recognized as really flaky.

\SetKwComment{Comment}{/* }{ */}

\begin{algorithm}[t!]
\small
\caption{FlaRe}\label{alg:flare}

  \SetKwInOut{Input}{inputs}
  \SetKwInOut{Output}{output}
  \SetKwProg{FlaRe}{FlaRe}{}{}

  \FlaRe{$(testcases,simData,tConditions,simModel)$}{
    \Input{testcases: Test Cases,
    \newline simData: Simulation Data,
    \newline tConditions: Test Conditions,
    \newline simModel: Simulation Model of CPS that have no real failures recognized by {\sc HyTestTF}}
    
    \SetKwProg{try}{try}{:}{}
    \SetKwProg{catch}{catch}{:}{end}
    \Output{flakyTests: Flaky Test Cases,
    \newline flakyModes: Flaky Modes}

    [failures, $\sim$ ]\ =\ testCPS($testcases,simData,...$\label{alg:80}
    
    $tConditions,simModels$);\label{alg:81}

    $modes =  getPotFlakyModes(tConditions)$; \label{alg:811} 
    
    \ForEach{test case $test_i \in testcases$}{ \label{alg:82}

        $startMode $ = extractStartState($test_i$);\label{alg:821}
        
        \uIf{$failures_i$}{\label{alg:83}
            $modes_{startMode}.append(``failed'')$ \label{alg:84}
        }
        \uElse{
            $modes_{startMode}.append(``passed'')$ \label{alg:85}
        }
    }

    \ForEach{flaky mode $mode_j \in modes$}{ \label{alg:86}

        \uIf{$mode_j$ contains ``failed'' and ``passed''}{\label{alg:87}

        add $mode_j$ to $flakyModes$; \label{alg:871}

            \ForEach{test case $test_i \in testcases$}{ \label{alg:88}

                \uIf{$test_i$'s start mode is $mode_j$}{\label{alg:89}
                
                    add $test_i$ to $flakyTests$;\label{alg:90}
                }
            
            }
            
        }
    }
    
    \KwRet{$flakyTests, flakyModes$}\;
  }

\end{algorithm}

\vspace*{-6pt}
\subsection{{\sc FlaRe}: Flaky Test Case Recognition}
\label{sec:flaky}

Once the faults have been fixed and {\sc HyTestFT} does not reveal real failures on the fixed model, it is time to determine the real flaky test cases. It is important to reveal real flaky test cases because every real flaky test case is the consequence of at least one failing and one passing flaky condition, starting from the same mode. When there is no real failure in the CPS, these flaky conditions are the result of incidents within the CPS or in the environment that the CPS is executing. Recognizing these flaky conditions can provide insights for CPS developers, allowing them to take preventive measure against those incidents. 

Algorithm \ref{alg:flare} shows our proposed approach, {\sc FlaRe}. {\sc FlaRe} receives test cases that are generated by {\sc HyTest}, test conditions, and simulation data as inputs and returns real flaky test cases and real flaky modes. 

First, in lines \ref{alg:80}-\ref{alg:81}, {\sc FlaRe} calls testCPS to test the corrected simulation model of the CPS using the test cases generated by {\sc HyTest}. Next, on line \ref{alg:811}, using the test conditions, which are the conditions on edges in the vstest condition graph, it finds all the modes that are likely flaky. {\sc FlaRe} marks a mode as a potentially flaky mode if one of the following cases is true.

\begin{itemize}
    \item If the mode has at least one outgoing edge to a failing mode and and at least one outgoing edge or edge to a passing  mode.

    \item If the mode has at least one outgoing edge to a failing mode and and at least one outgoing edge or edge to an acceptable mode.

    \item If the mode has at least one outgoing edge to a passing mode and and at least one outgoing edge or edge to an acceptable mode.

\end{itemize}

In the second and third cases, if the mode has an outgoing edge to an acceptable mode, then it may have a path to a passing mode or a failing mode or both; Hence, {\sc FlaRe} considers the mode as a potentially flaky mode. 

Then for each test case, it extracts the start mode (lines \ref{alg:82}-\ref{alg:821}). If the test case has failed/passed the test case, {\sc FlaRe} appends a ``failed''/``passed'' to its start mode to show that the CPS has failed/passed the test case starting from this mode (lines \ref{alg:83}-\ref{alg:85}). In the next step, lines \ref{alg:86}-\ref{alg:87}, {\sc FlaRe} looks for flaky modes. If a mode has a ``passed'' and ``failed'' it means the CPS has passed and failed the test case starting from this mode, so the mode is flaky and {\sc FlaRe} adds it to the set of real flaky modes (line \ref{alg:871}). Each test case that has a flaky start mode is a real flaky test case and is added to the set of real flaky test cases (lines \ref{alg:88}-\ref{alg:90}). Finally, {\sc FlaRe} returns the real flaky test cases and flaky modes. Since the test cases include their related test conditions and expected test verdicts, {\sc FlaRe}, implicitly, returns failing and passing flaky conditions as well.

\vspace*{-6pt}
\subsection{Implementation Details}

We used the {\sc HyTest}'s implementation described in our previous work \cite{SadriHyTest} and integrated the flakiness inspection module into it. Also, I implemented {\sc TReVa} and {\sc FlaRe} using Java 1.8.0 and Matlab R2022. 
Condition graph generation and test condition generation algorithms 
were implemented using Java and the rest of the implementation is in Matlab.
The Java code takes a hybrid model and 
other inputs, parses them, and generates a
condition graph and test conditions. 
The Matlab code obtains the test conditions and 
the rest of the inputs, generates and selects 
test cases, executes these test cases on the Simulink 
model of the CPS, returns and validate the test results and flaky test cases, modes, and conditions.

\section{Empirical Study}
\label{sec:experiment}

\noindent 
To evaluate {\sc HyTestTF}, with a particular focus
on the performance of {\sc TReVa} and {\sc FlaRe}, and the flakiness 
inspection performed by {\sc HyTestTF}`s test oracle, we conducted 
an empirical study that focused on the following questions:

\vspace*{6pt}
\noindent{\bf RQ1}: How effective is {\sc HyTestTF} at recognizing real failures, real passes, and real flaky failures?

\vspace*{6pt}
\noindent{\bf RQ2}: How effective is {\sc HyTestTF} at revealing distinct faults?\vspace*{6pt}

\noindent{\bf RQ3}: How effective is {\sc HyTestTF} at revealing real flaky modes?
\vspace*{6pt}

RQ1 is important because the correct recognition of real failures and real flaky failures can reduce fault localization and debugging efforts and consequently testing costs, by allowing test engineers to focus on actual failures and the faults that cause them, and not be distracted by the results of flaky test cases. 
Test cases that lead to flaky executions, having been identified, can also be repaired, rendering
them useful, or discarded, ensuring that they do not continue to be a distraction when test suites
are used later to test the system at different levels.
Finally, real passes are useful in that they help establish
whether the CPS has been designed and operates in accordance with its specifications.

While differentiating test cases that are flaky from those that are not is important, investigating whether real test failures reveal real faults within the CPS, and whether ignoring flaky test failures does not result in missing faults, is of higher importance. RQ2 helps us assess this; by assessing the numbers of distinct faults in the CPS that are related to the real failures it provides insights into how well {\sc HyTestTF} does at detecting faults in the CPS.

When faults in the CPS are fixed and {\sc HyTestTF} exhibits 
no more real failures, it is important to investigate whether the CPS 
can still exhibit flaky behavior. Given information on flaky modes, 
a CPS developer can implement measures to prevent 
the CPS from failing due to flaky conditions. 
RQ3 helps us determine the extent to which this could happen.

\subsection{Objects of Study}
\label{sec:objectsOfStudy}

To conduct our study we require CPSs. 
Just like {\sc HyTest}, {\sc HyTestTF} performs testing 
at the MiL level, and requires simulation models of the systems being tested. In this study, we used the same set of CPSs used to evaluate {\sc HyTest},
for which simulation models are available.\footnote{The
CPSs we selected are all available via links provided in the
relevant monographs, all of which are cited in this subsection.}
We provide a  description of those CPSs here. 
Table \ref{tab:objectsofstudy} provides data on the selected CPSs. 
The second column shows the number of blocks in the Simulink 
model of each CPS and the third column shows the number of 
modes in their hybrid models.

``Cruise Control'' \cite{CruiseControl}
monitors the speed of a vehicle through 
sensors, increases and decreases the vehicle's 
speed to match a set speed, and maintains that speed. 
``Inverted Pendulum'' is the system \cite{InvertedPendulum} 
introduced in Section \ref{subsec:CPS} and used 
to illustrate the operation of {\sc HyTest}. 
``Hexapod'' \cite{KhazaeeArticle} is an 
autonomous legged robot that is able to move around its 
environment with high flexibility and stability;
it can reach a preset goal and avoid preset 
obstacles on its way to the goal. 
``Rooms and Heaters'' \cite{FehnkerInPro} 
controls the transfer of moving heaters among 
adjacent rooms to maintain the temperature of 
the rooms within a desired range. 
Finally, ``Automatic Transmission'' controls an 
automobile's speed and engine rpm~\cite{ZhaoArticle}.

The simulation models for all five of these CPSs
are (hybrid) Simulink models, and they contain a wide range of 
block types including Integrator, Merge, Add, Transfer 
Function, If, State Chart, Relational and Logical Operators.
These Simulink models have each been 
designed using different types of controllers; this 
allows us to examine whether {\sc HyTest} can generate 
test cases for CPS's with different controllers.
For Inverted Pendulum we retrieved a hybrid 
model from \cite{INHybrid} and for Cruise Control 
we followed the description provided in 
\cite{CruiseControl} to recreate its model. 
The hybrid model for Hexapod is from \cite{KhazaeeArticle}. 
The hybrid models for Rooms and Heaters and Automatic 
Transmission are from \cite{AutomaticTransmissionHM1}
and \cite{AutomaticTransmissionHM2}. 

\begin{table}[t!]
\renewcommand\thetable{1}
\begin{center}    
\caption{ Objects of Study} 
\label{tab:objectsofstudy}
\vspace*{-6pt}
\resizebox{\columnwidth}{!}{\begin{tabular}{| c | c | c |}
\hline
& \# of Simulink  &  \# of Modes in 
\\
& Blocks & Hybrid Model\\
\hline
Cruise Control & 16 & 2 

\\  
\hline

Inverted Pendulum & 11 & 3 
\\ 
\hline

Hexapod & 348 & 5 
\\
\hline
Rooms and Heaters & 24 & 6
\\
\hline

Automatic Transmission& 14 & 4
\\ 
\hline
\end{tabular}
}
\end{center}
\vspace*{-10pt}
\end{table}

To answer our research questions we required 
information on fault detection and flakiness recognition for our object CPSs. 
Unfortunately, no faults have been reported for these systems, and finding 
events and conditions that cause flakiness in our CPSs was not feasible.
Thus, for the fault detection step of our study, we used the faulty models that were generated for use in evaluating {\sc HyTest} \cite{SadriHyTest}. These faulty models are models of CPSs that are mutated using a comprehensive list of Simulink fault patterns based on experiences reported by Delphi Engineers and from their review of the literature on CPSs~\cite{MatinnejadInPro}.

We also required ``flakiness'' in our test cases.
Although there are several techniques for injecting flakiness into test cases \cite{ChenInPro,CordyInPro,HabchiMisc} and some collections of real world flaky test cases exist \cite{GruberInPro,LamInPro1}, they all involve flakiness software systems that are written in specific programming languages such as Java or JavaScript, and do not involve simulation models. Furthermore, the flakiness involved is limited to known/identified sources of flakiness in those types of systems, and this omits other sources of flakiness. Therefore, we chose a different approach to reproduce the effects of flakiness: we injected random noises into simulation  models of our CPSs.

To reproduce the effects of flakiness, we used the original (unmutated) simulation models of the CPSs and added a noise block to the feedback loop of each CPS model to inject noise into the feedback values.
To inject noise, for each CPS, we randomly selected a random number $n$ that is used to randomly select $n$ output variables from the list of the CPS's outputs. Then for each randomly selected output variable, we selected a random value and added it to the value of that CPS's output variable at a randomly selected time. Because the noise value could be negative or positive, the addition is an algebraic sum. 
To determine how many noisy models to create, we measured the time required to test the faulty models and used that to guide the creation and testing of noisy models; this allowed us to conduct the study in a feasible amount of time. For example, the time required to test the faulty models of the Inverted Pendulum was around one minute, so we generated and tested as many noisy models as possible in one minute.

Tables~\ref{tab:TReVaCC}~-~\ref{tab:TReVaAT}, which are
presented in Section \ref{sec:results}, provide statistics on the number of faulty models (second column) and the number of noisy models (third column) obtained through this process for each of our objects of study.

\vspace*{-6pt}
\subsection{Variables and Measures}
\label{subsec:vars}

The independent variable in our study is the testing technique utilized.
{\sc HyTestTF} is one such technique, and as a baseline for comparison 
we chose {\sc HyTest}. {\sc HyTest} is a logical choice because it is the
only other extent technique for testing CPSs using simulation models.
Choosing {\sc HyTest} as a baseline allows us to assess the extent to which
{\sc HyTestTF} succeeds in revealing useful information related to test flakiness,
when considering research questions RQ1 and RQ2.

To perform the evaluation, we used the test results obtained by applying 
{\sc HyTest} to faulty models of our subject CPSs in our previous study of {\sc HyTest}. 
Then we tested the noisy models of all the CPSs that we generated as described above. 

Our dependent variables are the number of real failures, the number of real passes, the number of real flaky failures, the number of distinct faults that the real failures reveal, and the number of flaky modes.
These are measured to evaluate the effectiveness of {\sc HyTestTF} in validating test results. 
To measure them we wrote a script that counted the number of failures, passes, and 
distinct faults revealed by {\sc HyTest} and {\sc HyTestTF} on both faulty and noisy models. 
We wrote a second script to count the number of real flaky test failures 
and real flaky modes recognized by {\sc HyTestTF}. 

%

To evaluate the effectiveness of {\sc HyTestTF} in recognizing real flaky modes, 
we measured the number of modes that it recognized as such. 

\vspace*{-6pt}
\subsection{Study Operation}
\label{sec:study}

In practice, a user (test engineer, control designer, or CPS developer) of {\sc HyTestTF} would run it on a simulation model or CPS they designed to retrieve the validation results along with information on real flaky modes, real flaky test cases, and real flaky conditions. To evaluate the effectiveness of {\sc HyTestTF} in the context of this study, however, we used a three-step process applied to the faulty (mutated) and noisy models that were created for use in our study. The three-step process allows us to more easily capture the values of the dependent variables that we are concerned with.

In the first step of our process, for each of the five objects of study, {\sc HyTestTF} uses the same approach used by {\sc HyTest} to generate test cases and test faulty models of the CPS. {\sc HyTestTF}, like {\sc HyTest}, puts the CPS in different initial states by setting the model's initial values to the test inputs, providing the initial mode for the test oracle, and executing the faulty models. We record the number of failures, potentially flaky failures, and potential passes revealed by this step. Then, we run {\sc TReVa} to validate {\sc HyTestTF}'s test results on faulty models and we capture the numbers of real failures, real flaky test cases, real passes, and distinct faults (mutations) that were revealed by real failures. 

In the second step of our process, we generate noisy models (as discussed before) and repeat what we did on the faulty models in the first step of the process. Because {\sc HyTestT} does not recognize any potential flaky failures on some sets of noisy models when it is testing CPSs, we executed this second step 10 times. We did this in order to make sure we had noisy models that cover cases in which {\sc TReVa} did and did not recognize potential flaky failures. In this step, we also measured one more parameter -- the number of generated noisy models in each round of execution.  At this point of the study we have the results on how well {\sc HyTestTF} has done in providing validated test results.

In the third step of the process, we evaluate {\sc FlaRe}'s ability to recognize flaky modes. We use test cases generated by {\sc HyTestTF} in testing the original (unmutated) simulation models of the CPSs to conduct this step. We measure the number of real flaky modes that {\sc FlaRe} recognizes. 

Finally, to assess the accuracy of the results, we compared the results obtained from {\sc HyTestTF} with those obtained by a manual assessment method, through the following process.

To validate measures for the variables from {\sc HyTestTF} where {\sc TReVa} was concerned, we inspected the test results from the test oracle and the validated results that {\sc HyTestTF} provided. If a failure occurred on faulty models both times, we recognized it as a real failure. To obtain measures for other variables we implemented code to inspect the results. The code checked whether the output values fit any failing conditions, and if they did, we marked the test execution as a potential failure. If the output values did not fit any failing conditions and the output values fit the goal of the CPS at the end of the simulation, we marked the test execution as a potential pass. When test executions were marked differently across executions we counted them as real flaky test cases. When test executions were marked the same across executions we counted them as real passes or real failures.

To obtain measures for the variables from {\sc HyTestTF} where {\sc FlaRe} was concerned, we looked in the condition graph for modes that had at least one outgoing passing transition and at least one outgoing failing transition. Then, we used the test cases related to those transitions to test the CPS and checked the results. If the CPS passed the test using the passing test case and failed the test using the failing test case, then we marked the mode as a real flaky mode and marked those test cases as flaky test cases. 

The results of our manual validation matched the results returned by {\sc HyTestTF}  for all of our subject CPSs.

\vspace*{-6pt}
\subsection{Threats to Validity}

{\em External validity} threats concern the generalization of our findings.
As objects of study we selected five CPSs, so our results pertain
to those CPSs and may not generalize beyond them.
To answer our research questions we relied on the insertion of 
mutations into Simulink models. Our mutations are based on those defined 
in earlier research in which they were derived based on experiences
from Delphi Engineers and other sources~\cite{MatinnejadInPro}, so they
have some relevance to natural faults; nevertheless,
results obtained using these may not match the results 
that could be obtained on natural faults occurring in practice.

The hybrid models obtained for our objects
represent only five specific models.
However, we did verify the simulation models against the 
hybrid models and specifications and made sure both models 
behaved in accordance with their corresponding specifications. 

{\em Internal validity} threats concern uncontrolled 
factors that may have affected our results.
Errors in our implementation of {\sc HyTestTF}
and the manual assessment method we used to
verify the correctness of that implementation
could do so.

{\em Construct validity} threats concern our metrics and measures, and 
how well they actually capture the effects we are trying to measure. 
The metrics we capture concern numbers of real failures, real passes,
real flaky failures, distinct faults revealed, and flaky modes.
The extent to which these metrics reflect potential efficiency
gains in the underlying testing and debugging process is not directly examined.

\vspace*{-6pt}
\subsection{Results}
\label{sec:results}

Tables \ref{tab:TReVaCC}, \ref{tab:TReVaIN}, \ref{tab:TReVaHX}, 
\ref{tab:TReVaRH}, and \ref{tab:TReVaAT} show the results of our study
for each of the five object CPSs in turn.
In these tables, the number of test cases that {\sc HyTest} and {\sc HyTestTF} 
have generated and used to test those CPSs is noted in the caption. 
In a given table, each row shows the results of one of the ten rounds that were
conducted to generate noisy models for that object CPS, test them using 
{\sc HyTestTF}, and run {\sc TreVa} to validate the test results (rounds are numbered in Column~1).
Column 2 gives the number of faulty models created for the CPS, and 
Column 3 gives the number of noisy models generated in each round.  
Columns 4, 6, and 8 show the numbers of real failures, real passes, 
and real flaky runs recognized by {\sc HyTestTF},
while Column 5 and 7 show the numbers of real failures and real
passes recognized by {\sc HyTest}. 
Column 9 shows the percentage of test runs that are real flaky runs. 
Columns 10 and 11 show the number of distinct faults, i.e., the number 
of mutations plus the number of real faults (if any) found in the original 
model\footnote{ As we reported in our previous work \cite{SadriHyTest}, 
{\sc HyTest} revealed some actual failures while we were executing 
our implementation of it using the original models, i.e. models that were not mutated. 
Those failures were related to one fault in each of the CPSs.} 
that were detected by {\sc HyTestTF} and {\sc HyTest}, respectively.

\begin{table*}[p]
\begin{center}
\renewcommand\thetable{2.1}
\footnotesize
\caption{ Results of testing Cruise Control using 11 test cases} 
\label{tab:TReVaCC}
\begin{tabular}{| c | c | c | c | c | c | c | c | c | c | c |}

\hline Round & Faulty  & Noisy  & Failures & Failures & Passes & Passes & Flaky & Flaky Runs  & Distinct Faults & Distinct Faults  \\ 

&Models & Models & ({\sc HyTestTF}) & ({\sc HyTest}) & ({\sc HyTestTF}) & ({\sc HyTest}) & Runs & Percentage & ({\sc HyTestTF}) & ({\sc HyTest}) \\ 

\hline 1 & 26 & 16 & 148 & 248 & 144 & 214 & 170 & 36.8\% & 27 & 27 \\

\hline 2 & 26 & 16 & 150 & 278 & 115 & 184 & 197 & 42.6\% & 27 & 27 \\

\hline 3 & 26 & 15 & 153 & 291 & 97 & 160 & 201 & 44.6\% & 27 & 27 \\

\hline 4 & 26 & 16 & 156 & 262 & 137 & 200 & 169 & 36.6\% & 27 & 27 \\

\hline 5 & 26 & 16 & 149 & 254 & 140 & 208 & 173 & 37.5\% & 27 & 27 \\

\hline 6 & 26 & 18 & 162 & 308 & 113 & 176 & 209 & 43.2\% & 27 & 27 \\

\hline 7 & 26 & 18 & 146 & 286 & 122 & 198 & 216 & 44.6\% & 27 & 27 \\

\hline 8 & 26 & 18 & 162 & 277 & 144 & 207 & 178 & 36.8\% & 27 & 27 \\

\hline 9 & 26 & 18 & 162 & 278 & 143 & 206 & 179 & 37.0\% & 27 & 27 \\

\hline 10 & 26 & 17 & 146 & 279 & 120 & 194 & 207 & 43.8\% & 27 & 27 \\

\hline

 \end{tabular}
 \vspace*{-12pt}
 \end{center}
\end{table*}

\begin{table*}[p]
\begin{center}
\renewcommand\thetable{2.2}
\footnotesize
\caption{ Results of testing Inverted Pendulum using 13 test cases} 
\label{tab:TReVaIN}
\begin{tabular}{| c | c | c | c | c | c | c | c | c | c | c |}

\hline Round & Faulty  & Noisy  & Failures & Failures & Passes & Passes & Flaky & Flaky Runs  & Distinct Faults & Distinct Faults  \\ 

&Models & Models & ({\sc HyTestTF}) & ({\sc HyTest}) & ({\sc HyTestTF}) & ({\sc HyTest}) & Runs & Percentage & ({\sc HyTestTF}) & ({\sc HyTest}) \\ 

\hline 1 & 20 & 22 & 176 & 348 & 157 & 198 & 213 & 39.0\% & 21 & 21 \\

\hline 2 & 20 & 24 & 182 & 329 & 201 & 243 & 189 & 33.0\% & 21 & 21 \\

\hline 3 & 20 & 19 & 180 & 282 & 195 & 225 & 132 & 26.0\% & 21 & 21 \\

\hline 4 & 20 & 20 & 186 & 295 & 197 & 225 & 137 & 26.4\% & 21 & 21 \\

\hline 5 & 20 & 21 & 175 & 326 & 169 & 207 & 189 & 35.6\% & 21 & 21 \\

\hline 6 & 20 & 23 & 175 & 334 & 180 & 225 & 204 & 36.5\% & 21 & 21 \\

\hline 7 & 20 & 23 & 194 & 352 & 157 & 207 & 208 & 37.2\% & 21 & 21 \\

\hline 8 & 20 & 23 & 165 & 300 & 205 & 259 & 189 & 33.8\% & 21 & 21 \\

\hline 9 & 20 & 23 & 188 & 370 & 143 & 189 & 228 & 40.8\% & 21 & 21 \\

\hline 10 & 20 & 23 & 196 & 325 & 204 & 234 & 159 & 28.4\% & 21 & 21 \\

\hline
 \end{tabular}
 \vspace*{-12pt}
 \end{center}
\end{table*}

\begin{table*}[p]
\begin{center}
\renewcommand\thetable{2.3}
\footnotesize
\caption{ Results of testing Hexapod using 20 test cases} 
\label{tab:TReVaHX}
\begin{tabular}{| c | c | c | c | c | c | c | c | c | c | c |}

\hline Round & Faulty  & Noisy  & Failures & Failures & Passes & Passes & Flaky & Flaky Runs  & Distinct Faults & Distinct Faults  \\ 

&Models & Models & ({\sc HyTestTF}) & ({\sc HyTest}) & ({\sc HyTestTF}) & ({\sc HyTest}) & Runs & Percentage & ({\sc HyTestTF}) & ({\sc HyTest}) \\ 

\hline 1 & 191 & 26 & 1202 & 2073 & 1335 & 2267 & 1803 & 41.5\% & 192 & 192 \\

\hline 2 & 191 & 26 & 1202 & 2077 & 1331 & 2263 & 1807 & 41.6\% & 192 & 192 \\

\hline 3 & 191 & 26 & 1135 & 2088 & 1260 & 2252 & 1945 & 44.8\% & 192 & 192 \\

\hline 4 & 191 & 24 & 1149 & 2073 & 1258 & 2227 & 1893 & 44.0\% & 192 & 192 \\

\hline 5 & 191 & 18 & 1136 & 2042 & 1175 & 2138 & 1869 & 44.7\% & 192 & 192 \\

\hline 6 & 191 & 18 & 1108 & 2033 & 1173 & 2147 & 1899 & 45.4\% & 192 & 192 \\

\hline 7 & 191 & 20 & 1118 & 2048 & 1194 & 2172 & 1908 & 45.2\% & 192 & 192 \\

\hline 8 & 191 & 26 & 1186 & 2082 & 1312 & 2258 & 1842 & 42.4\% & 192 & 192 \\

\hline 9 & 191 & 23 & 1111 & 2086 & 1175 & 2194 & 1994 & 46.6\% & 192 & 192 \\

\hline 10 & 191 & 29 & 1178 & 2125 & 1268 & 2275 & 1954 & 44.4\% & 192 & 192 \\

\hline
 \end{tabular}
 \vspace*{-12pt}
 \end{center}
\end{table*}

\begin{table*}[p]
\begin{center}
\renewcommand\thetable{2.4}
\footnotesize
\caption{ Results of testing Rooms and Heaters using 43 test cases} 
\label{tab:TReVaRH}
\begin{tabular}{| c | c | c | c | c | c | c | c | c | c | c |}

\hline Round & Faulty  & Noisy  & Failures & Failures & Passes & Passes & Flaky & Flaky Runs  & Distinct Faults & Distinct Faults  \\ 

& Models & Models & ({\sc HyTestTF}) & ({\sc HyTest}) & ({\sc HyTestTF}) & ({\sc HyTest}) & Runs & Percentage & ({\sc HyTestTF}) & ({\sc HyTest}) \\  

\hline 1 & 56 & 10 & 562 & 1092 & 1024 & 1746 & 1252 & 44.1\% & 57 & 57 \\

\hline 2 & 56 & 9 & 556 & 1085 & 993 & 1710 & 1246 & 44.6\% & 57 & 57 \\

\hline 3 & 56 & 11 & 627 & 1081 & 1131 & 1800 & 1123 & 39.0\%  & 57 & 57 \\

\hline 4 & 56 & 9 & 613 & 1067 & 1059 & 1728 & 1123 & 40.2\%  & 57 & 57 \\

\hline 5 & 56 & 9 & 556 & 1085 & 993 & 1710 & 1246 & 44.6\%  & 57 & 57 \\

\hline 6 & 56 & 9 & 613 & 1067 & 1059 & 1728 & 1123 & 40.2\%  & 57 & 57 \\

\hline 7 & 56 & 9 & 559 & 1094 & 997 & 1701 & 1239 & 44.3\%  & 57 & 57 \\

\hline 8 & 56 & 9 & 559 & 1076 & 1001 & 1719 & 1235 & 44.2\%  & 57 & 57 \\

\hline 9 & 56 & 10 & 560 & 1083 & 1026 & 1755 & 1252 & 44.1\%  & 57 & 57 \\

\hline 10 & 56 & 7 & 559 & 1062 & 945 & 1647 & 1205 & 44.5\%  & 57 & 57 \\

\hline
 \end{tabular}
 \vspace*{-12pt}
 \end{center}
\end{table*}

\begin{table*}[t]
\begin{center}
\renewcommand\thetable{2.5}
\footnotesize
\caption{ Results of testing Automatic Transmission using 25 test cases} 
\label{tab:TReVaAT}
\begin{tabular}{| c | c | c | c | c | c | c | c | c | c | c |}

\hline Round & Faulty  & Noisy  & Failures & Failures & Passes & Passes & Flaky & Flaky Runs  & Distinct Faults & Distinct Faults  \\ 

&Models & Models & ({\sc HyTestTF}) & ({\sc HyTest}) & ({\sc HyTestTF}) & ({\sc HyTest}) & Runs & Percentage & ({\sc HyTestTF}) & ({\sc HyTest}) \\ 

\hline 1 & 66 & 51 & 680 & 2105 & 665 & 820 & 1580 & 54.0\% & 66 & 67 \\

\hline 2 & 66 & 16 & 439 & 1430 & 496 & 620 & 1115 & 54.4\% & 66 & 67 \\

\hline 3 & 66 & 47 & 792 & 2055 & 606 & 770 & 1427 & 50.5\% & 66 & 67 \\

\hline 4 & 66 & 37 & 571 & 1825 & 599 & 750 & 1405 & 54.6\% & 66 & 67 \\

\hline 5 & 66 & 9 & 339 & 1275 & 487 & 600 & 1049 & 56.0\% & 66 & 67 \\

\hline 6 & 66 & 42 & 607 & 1910 & 636 & 790 & 1457 & 54.0\% & 66 & 67 \\

\hline 7 & 66 & 38 & 580 & 1840 & 611 & 760 & 1409 & 54.2\% & 66 & 67 \\

\hline 8 & 66 & 44 & 560 & 1930 & 647 & 820 & 1543 & 56.1\% & 66 & 67 \\

\hline 9 & 66 & 44 & 607 & 1940 & 648 & 810 & 1495 & 54.4\% & 66 & 67 \\

\hline 10 & 66 & 43 & 662 & 1945 & 613 & 780 & 1450 & 53.2\% & 66 & 67 \\

\hline
\end{tabular}
\vspace*{-12pt}
\end{center}
\end{table*}

\begin{table*}[t]
\begin{center}
\renewcommand\thetable{3}
\footnotesize
\caption{ Descriptive Statistics of Flaky Run Percentages} 
\label{tab:stats}
\begin{tabular}{| c | c | c | c | c | c | }

\hline & Min-Max & Range & Mean  & Median  & Standard Deviation \\

\hline  Cruise Control & 36.6\% - 44.6\% & 8.0\% & 40.3\% & 40.0\% & 3.5\% \\ 

\hline Inverted Pendulum & 26.0\% - 40.8\% & 14.8\% & 33.7\% & 34.6\% &  4.9\% \\ 

\hline Hexapod & 41.5\% - 46.6\% & 5.1\% & 44.1\% & 44.6\%  & 1.6\% \\ 

\hline Rooms and Heaters & 39.0\% - 44.6\% & 5.6\% & 43.5\% & 44.1\% & 2.2\% \\ 

\hline Automatic Transmission & 50.5\% - 56.1\% & 5.6\% & 54.1\% & 54.3\% & 1.5\% \\

\hline
\end{tabular}
\vspace*{-12pt}
\end{center}
\end{table*}

We now consider research questions RQ1 and RQ2.
We begin with the data for the Cruise Control CPS, shown in Table \ref{tab:TReVaCC}.
As Table \ref{tab:TReVaCC} shows, during the first round of the study, 
26 faulty and 16 noisy models were tested by {\sc HyTest} and {\sc HyTestTF}; 
the results of testing were validated by {\sc HyTestTF}. 
Because the number of test cases used was 11, and each model was 
tested using all of the test cases, the total number of test 
runs for both {\sc HyTest} and {\sc HyTestTF} is $11 \times (26 + 16) = 462$. 
Among these test runs {\sc HyTest} recognized 248 failures and  
214  passes while {\sc HyTestTF} recognized  148  real failures, 
144  real passes, and 170 flaky test runs. 
For both {\sc HyTest} and {\sc HyTestTF}, these numbers 
add up to 462, which is the total number of test runs. 
This shows that {\sc HyTestTF} has validated the results 
of all test runs for Cruise Control during the first round. 

Considering the results across all of Tables \ref{tab:TReVaCC} to \ref{tab:TReVaAT}, a substantial portion of the test runs (from 26\% to 56.1\%) are flaky across all of our subject CPSs. 
For Cruise Control the mean percentage of flaky tests is 40.3\%. The median value of 40.0\% is very close to the mean and indicates a fairly symmetric distribution of flaky test percentages, with no extreme outliers skewing the data significantly. This is true for all other CPSs except for Inverted Pendulum, where the median (34.6\%) is nearly 1\% apart from mean (33.7\%).
These levels of flakiness can undermine the overall effectiveness and efficiency of the testing process. Also, the data shows that for all CPSs, as for Cruise Control,
{\sc HyTestTF} validated the results of all test runs during the first round. 

For another look at the data pertaining to flaky test runs,
Table \ref{tab:stats} provides descriptive statistics for the flaky run percentages, i.e., data in Column 9 in Tables \ref{tab:TReVaCC} to \ref{tab:TReVaAT}, for all of our CPSs. Column 1 names the CPSs. Columns 2 through 5 show the minimum and maximum values, range, mean, median, and standard deviation of the flaky runs, respectively. For example, the first row of the table provides data for Cruise Control. As the Table \ref{tab:stats} shows, for Cruise Control, the minimum percentage of flaky test runs was 36.6\%, so at best, 63.4\% of the test runs were not flaky. The maximum percentage was 44.6\%, so at best, of the 55.4\% of the test runs were not flaky. Thus, the validity of the test results varied by 8\% across different runs, with a standard deviation of 3.5\%. This 8\% range together with a standard deviation of 3.5\% indicate a moderate variability in the percentage of flaky test runs across different test runs. For Inverted Pendulum the range is 14.8\% and the standard deviation is 4.9\%. Compared to Cruise Control,  it shows a higher variability around the mean. For Hexapod, Rooms and Heaters, and Automatic Transmission this variability is around 5\%, which is less than the range for the first two CPSs. Also, for these systems the standard deviation is less than for the first two CPSs. This suggests less variability in the percentage of flaky test runs in these CPSs, as well.  

\begin{figure}[b!]
\centering
  \includegraphics[width=\columnwidth]{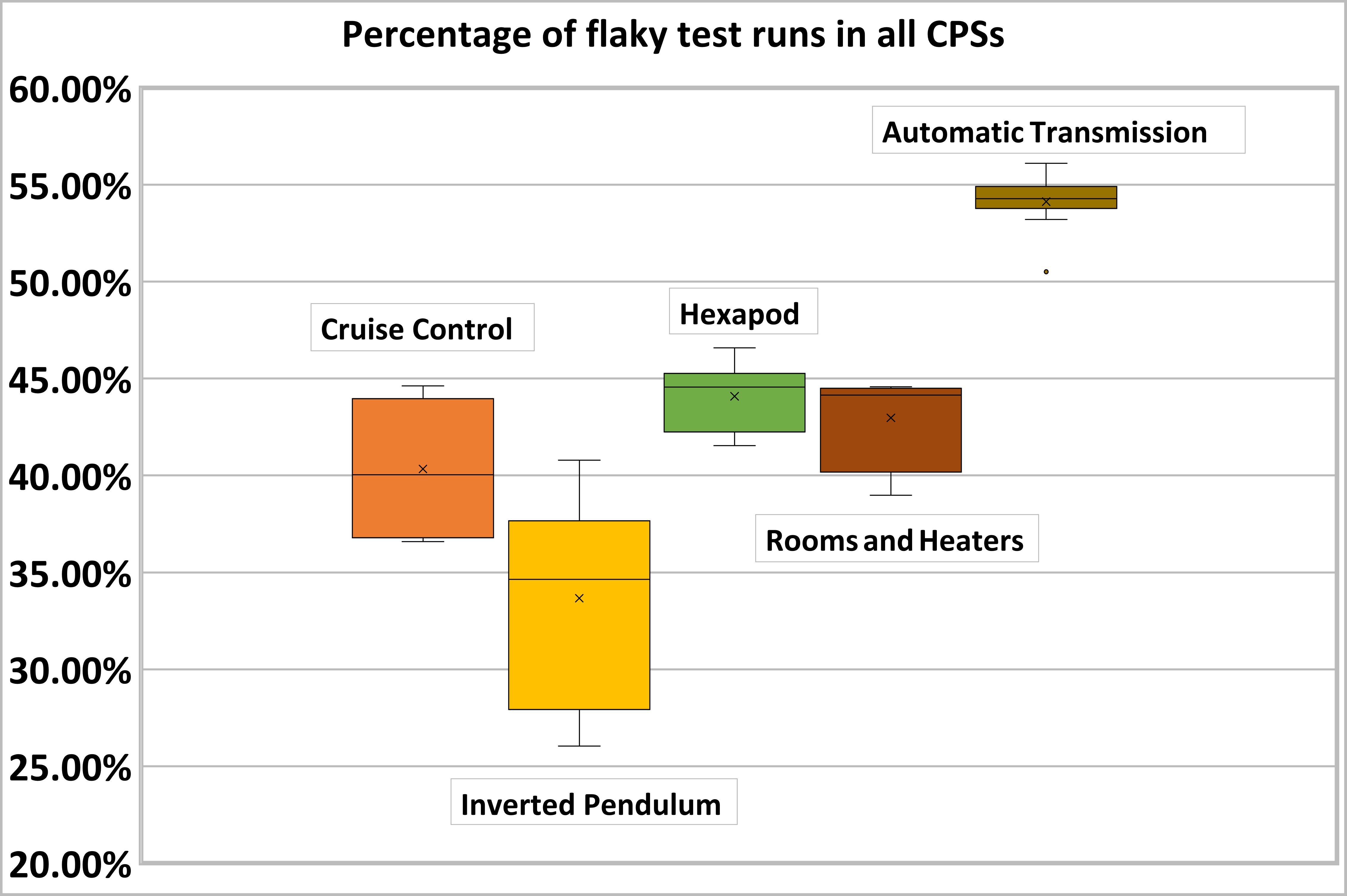}
  \caption{Percentage of flaky test runs in all CPSs}
  \label{fig:FlakyPercentage}
\vspace*{-12pt}
\end{figure}

Figure \ref{fig:FlakyPercentage} provides a graphical view of the data 
from Table \ref{tab:stats}, using boxplots to depict the ranges of
percentages of flaky test runs observed on the five CPSs studied.
As the boxplots show, the largest variability in results occurred
on Inverted Pendulum, followed closely by Cruise Control.
The other three CPSs display substantially less variation in 
percentages of flaky test runs observed.

To answer RQ1, on all five CPSs, since {\sc HyTestTF} was able to differentiate real 
failures and passes from those that were flaky, {\sc HyTestTF} was more effective 
than {\sc HyTest} in recognizing real failures, real passes, and real flaky 
failures.

Next we turn to research question RQ2.
As noted earlier, when evaluating {\sc HyTestTF}, while it is important to 
differentiate real failures and passes from flaky failures or passes, 
it is also important to asses the effectiveness of the approach in revealing 
faults.\footnote{By ``revealing faults'' we mean that the approach determines that a 
fault must exist; this does not mean that it localizes the fault}. 
To this end, we recorded the distinct faults that each failure could reveal.
Data in Columns 10 and 11 of Tables \ref{tab:TReVaCC} to \ref{tab:TReVaAT}
shows the number of distinct faults, 
which is the number of faulty models plus the number of real faults in 
the original model for each object of study, that {\sc HyTestTF} and {\sc HyTest} revealed. 
This data shows that discarding the real flaky failures did not reduce the
fault revealing ability of {\sc HyTestTF} on any CPS other than Automatic Transmission; 
across the other four CPSs {\sc HyTestTF} 
revealed as many faults that {\sc HyTest}, i.e., all the 
faults we had injected or recognized previously. 

For Automatic Transmission, we first confirmed that {\sc HyTestTF} could not reveal one of the faults, but after carefully and manually investigating the results, we noticed that by changing the conditions under which the Automatic Transmission was working, it alternately passed and failed the test, which means those test runs were real flaky ones. Although initially it seems that {\sc HyTestTF}, and specifically {\sc TReVa} has missed one of the faults, it is worth noting that after validating the test results, {\sc HyTestTF}, specifically {\sc FlaRe} recognizes real flaky modes, test cases, and conditions, and provides these data for the test engineer or system developer, so they can execute the CPS under each of the flaky failing conditions and localize their relevant faults. This means finally that {\sc HyTestTF} reveals all faults, no matter they are flaky faults or not.

Finally, we turn to research question RQ3.
Table \ref{tab:resFL} shows the results of our study for {\sc HyTestTF} 
(specifically for {\sc FlaRe}) for each of the CPSs. 
The first column of the table contains names of the CPSs. 
The second column shows the number of modes that are contained
in the condition graph generated by {\sc HyTestTF} for each CPS. 
The third column shows the number of potentially flaky modes 
that {\sc HyTestTF} has recognized for each CPS. 
The fourth column shows the number of real flaky modes that {\sc HyTestTF} recognized. 
Our manual validation process confirmed the correctness of these results. 
The results show that {\sc HyTestTF} was able to correctly recognize 
all of the flaky modes in all of the CPSs studied.

\begin{table}[t]
\begin{center}
\renewcommand\thetable{4}
\caption{Results for Flaky-Modes Recognition of {\sc HyTestTF}} 
\label{tab:resFL}

\begin{tabular}{| c | c | c | c |}

\hline  & Modes in Test  & Potential  &   Real   \\

 & Condition Graph & Flaky &  Flaky  \\ 

 & Graph & Modes &  Modes  \\ 

\hline  Cruise Control & 3 & 2  & 2 \\ 

\hline Inverted Pendulum  & 4 &  3 & 3 \\ 

\hline Hexapod & 6 & 5 &  5 \\ 

\hline Rooms and Heaters & 7 & 6 &  6 \\ 

\hline Automatic Transmission & 5 & 4 &  4 \\

\hline
 \end{tabular}
 \vspace*{-12pt}
 \end{center}
\end{table}

\section{Discussion}
\label{sec:discussion}

In this section, we discuss the results of our empirical study, 
and observations about the results.

\begin{table*}[!b]
\centering

\renewcommand\thetable{5}
\footnotesize
\caption{ Flaky Test Runs} 
\label{tab:FlakyTestRunsDis}
\begin{tabular}{|@{}c@{}||@{}c@{}||@{}c@{}||@{}c@{}||@{}c@{}||@{}c@{}@{}|}

\hline 

\begin{tabular}{c}
\\
\\
Round \\ \hline
    1 \\ \hline
    2 \\ \hline
    3 \\ \hline
    4 \\ \hline
    5 \\ \hline
    6 \\ \hline
    7 \\ \hline
    8 \\ \hline
    9 \\ \hline
    10 \\ 
    
\end{tabular}
  
   & \begin{tabular}{@{}c@{}} \scriptsize Cruise Control \\
       \hline
       \begin{tabular}{c|c|c}
 
        \scriptsize Flaky & \scriptsize Flaky & \scriptsize Flaky \\
        \scriptsize Runs & \scriptsize Failures & \scriptsize Passes\\ \hline
         170 & 58.8\% &  41.2\% \\ \hline
         197 & 65.0\% &  35.0\% \\  \hline
         201 & 68.7\% &  31.3\% \\ \hline
         169 & 62.7\% &  37.3\% \\ \hline
         173 & 60.7\% &  39.3\% \\ \hline
         209 & 69.9\% &  30.1\% \\ \hline
         216 & 64.8\% &  35.2\% \\ \hline
         178 & 64.6\% &  35.4\% \\ \hline
         179 & 64.8\% &  35.2\% \\ \hline
         207 & 64.3\% &  35.7\% \\ 
        \end{tabular}
    
    \end{tabular} 

    & \begin{tabular}{@{}c@{}} \scriptsize Inverted Pendulum \\
       \hline
       \begin{tabular}{c|c|c}
 
        \scriptsize Flaky & \scriptsize Flaky & \scriptsize Flaky \\
        \scriptsize Runs & \scriptsize Failures & \scriptsize Passes\\ \hline
         213 & 80.8\% &  19.2\% \\ \hline
         189 & 77.8\% &  22.2\% \\  \hline
         132 & 77.3\% &  22.7\% \\ \hline
         137 & 79.6\% &  20.4\% \\ \hline
         189 & 79.9\% &  20.1\% \\ \hline
         204 & 77.9\% &  22.1\% \\ \hline
         208 & 76.0\% &  24.0\% \\ \hline
         189 & 71.4\% &  28.6\% \\ \hline
         228 & 79.8\% &  20.1\% \\ \hline
         159 & 81.1\% &  18.9\% \\ 
        \end{tabular}
    
    \end{tabular} 

    & \begin{tabular}{@{}c@{}} \scriptsize Hexapod \\
       \hline
       \begin{tabular}{c|c|c}
 
        \scriptsize Flaky & \scriptsize Flaky & \scriptsize Flaky \\
        \scriptsize Runs & \scriptsize Failures & \scriptsize Passes\\ \hline
         1803 & 48.3\% &  51.7\% \\ \hline
         1807 & 48.4\% &  51.6\% \\  \hline
         1945 & 49.0\% &  51.0\% \\ \hline
         1893 & 48.8\% &  51.2\% \\ \hline
         1869 & 48.5\% &  51.5\% \\ \hline
         1899 & 48.7\% &  51.3\% \\ \hline
         1908 & 48.7\% &  51.3\% \\ \hline
         1842 & 48.6\% &  51.4\% \\ \hline
         1994 & 48.9\% &  51.1\% \\ \hline
         1954 & 48.5\% &  51.5\% \\  
        \end{tabular}
    
    \end{tabular} 

    & \begin{tabular}{@{}c@{}} \scriptsize Rooms and Heaters \\
       \hline
       \begin{tabular}{c|c|c}
 
        \scriptsize Flaky & \scriptsize Flaky & \scriptsize Flaky \\
        \scriptsize Runs & \scriptsize Failures & \scriptsize Passes\\ \hline
         1252 & 42.3\% &  57.7\% \\ \hline
         1246 & 42.5\% &  57.5\% \\  \hline
         1123 & 40.4\% &  59.6\% \\ \hline
         1123 & 40.4\% &  59.6\% \\ \hline
         1246 & 42.5\% &  57.5\% \\ \hline
         1123 & 40.4\% &  59.6\% \\ \hline
         1239 & 43.2\% &  56.8\% \\ \hline
         1235 & 41.9\% &  58.1\% \\ \hline
         1252 & 41.8\% &  58.2\% \\ \hline
         1205 & 41.7\% &  58.3\% \\ 
        \end{tabular}
    
    \end{tabular} 

    & \begin{tabular}{@{}c@{}} \scriptsize Automatic Transmission \\
       \hline
       \begin{tabular}{c|c|c}
 
        \scriptsize Flaky & \scriptsize Flaky & \scriptsize Flaky \\
        \scriptsize Runs & \scriptsize Failures & \scriptsize Passes\\ \hline
         1580 & 90.2\% &  9.8\% \\ \hline
         1115 & 88.9\% &  11.1\% \\  \hline
         1427 & 88.5\% &  11.5\% \\ \hline
         1405 & 89.3\% &  10.7\% \\ \hline
         1049 & 89.2\% &  10.8\% \\ \hline
         1457 & 89.4\% &  10.6\% \\ \hline
         1409 & 89.4\% &  10.6\% \\ \hline
         1543 & 88.8\% &  11.2\% \\ \hline
         1495 & 89.2\% &  10.8\% \\ \hline
         1450 & 88.5\% &  11.5\% \\ 
        \end{tabular}
    
    \end{tabular} 
\\
\hline 

\end{tabular}

\vspace*{-12pt}

\end{table*}

\subsection{Implications for Efficiency 
and Correctness}

Efficiency involves achieving a result with minimal extraneous or redundant effort \cite{Vocab}. In testing CPSs, this result includes generating test cases, executing the system using those test cases, detecting failures, localizing faults, and fixing them; hence, a testing techniques that is efficient accomplishes these things with minimal effort. Correctness, as an attribute of software, is the extent to which the software, documentation, or other items fulfill user needs and expectations, whether explicitly stated or not \cite{Vocab}. As noted earlier, the focus of RQ1 is on reducing the effort required for fault localization/debugging of CPSs and consequently reducing the effort required for testing CPSs, as well as providing insights on whether the CPS operates as it is desired by the end user or not. The former focuses on the efficiency of the test process, the latter on the CPSs correctness.

Regarding reduction of testing effort, real flaky failures involve test runs 
that failed during the testing phase but did not fail during the validation process.
When real flaky failures are not identified, and when the test engineer receives the test results on failures, they begin the debugging process to find the fault(s) that have caused the failures. When a failure is a flaky failure, since the flaky failing condition is unknown and the CPS randomly fails and passes the test, it can be difficult to debug the CPS under the same failing condition and find faults. Although there are several techniques that can recognize flaky test cases in software systems or can detect root causes of flakiness in CPSs, there are limitations for applying them to CPSs (as we discuss in Section \ref{sec:related}) that increases the chances of failing to recognize flaky test cases.

Regarding the issue of correctness, real flaky passes involve test runs that passed during the testing phase but did not pass during the validation process. Using the real passes, a CPS designer can obtain more accurate and reliable insights into whether the CPS they have designed conforms to the CPS specification, i.e., to the end user's expectations and needs. When real flaky passes are not identified, such insights are not reliable because it is unclear whether the CPS unconditionally responds to the end user's needs or just responds to their needs under some conditions and not under others.

To measure the number of real flaky failures in our study, we subtracted the number of real failures that {\sc HyTestTF} recognized from  the number of failures that {\sc HyTest} reported. Similarly, to measure the number of real flaky passes in our study, we subtracted the number of real passes that {\sc HyTestTF} recognized from the number of passes that {\sc HyTest} reported. Note that these two numbers add up to the total number of real flaky test runs. We report the results of this step in Table~\ref{tab:FlakyTestRunsDis}.

Column 1 in Table \ref{tab:FlakyTestRunsDis} shows the rounds of execution. The rest of the table has five sections, one for each object CPS. In each section, the first column shows the total number of real flaky test runs recognized by {\sc HyTestTF}, while the second and third columns show the percentages of real flaky failures and real flaky passes that made up those test runs, respectively. As an example, the first row in the first section shows that on Cruise Control, during the first round of execution, {\sc HyTestTF} recognized 170 real flaky runs, of which 58.8\% were real flaky failures and 41.2\% were real flaky passes. 

As the table shows, the minimum percentage of real flaky failures within a set of flaky runs is 40.0\%, while the maximum percentage is 90.2\%. By discarding this number of real flaky failures, the focus of fault localization and debugging will be on the real failures; Hence, the testing effort and costs will be reduced by using {\sc HyTestTF}. Also, the minimum percentage of real flaky passes is 9.8\% and the maximum is 59.6\%, i.e. around 10\% to 60\% of real flaky runs, over all CPSs, falsely show that the CPS works as it is expected while under a different condition it does not respond to the end user's needs.

\subsection{Faults Versus Design Problems}

As we noted in Section \ref{sec:results}, the number of 
distinct faults that were revealed by real failures was 
the sum of the number of faulty models we generated 
for the empirical study in our previous work \cite{SadriHyTest} 
and the number of faults that {\sc HyTest} could reveal in 
the original simulation models of the CPSs when we were 
testing the correctness of our implemented approach. 
Here it is worth noting that while a failure is an event where a system or system component 
fails to perform a required function within specified limits \cite{Vocab}, 
a fault is an actual defect in a system. 
One fault may cause zero or more failures and one failure 
may be the reason of one or more faults. 
Failures that {\sc HyTest} revealed in the original models
occurred because of a specific fault that was present in each of the subject CPSs: 
the CPS kept executing when they reached failing conditions. 
Given that none of our subject CPSs are supposed to continue 
running when they fail because doing so may put devices or 
humans under risk, we consider this a problem in the 
design of the CPSs and we count this as one fault.

\subsection{Insights about Flaky Behaviors}

Although the results of our study for {\sc FlaRe} do not 
demonstrate it, it is possible for the number of {\em potential} flaky 
modes recognized by {\sc FlaRe} to be greater than the number of real flaky modes it recognizes.
In this case, if {\sc FlaRe} does not mark a potentially flaky mode or 
test case as a real one, it does not necessarily mean that 
{\sc FlaRe} has missed them and they are false positives. 
It may be the case that based on the condition graph they are found to be
potentially flaky and the system designer/developer has already considered 
preventive measures in their CPS design. Such a case is a true negative, 
i.e., {\sc FlaRe} has correctly not inferred that the potentially flaky 
mode or test case is flaky.

Finally, when {\sc FlaRe} recognizes real flaky modes, it returns 
the test cases that cause flakiness starting from those modes. 
Since each test case that is generated by {\sc HyTestTF} contains 
the test conditions along with values that fit in those conditions, 
the test engineer can use those values to start the CPS from the 
flaky modes under those flaky conditions, run the CPS, and 
investigate the CPS' behavior, in order to find preventative 
measures by which those flaky failing conditions can be avoided.

\section{Related Work}
\label{sec:related}

Several techniques have been proposed 
for recognizing flaky test cases in software systems 
\cite{BellInPro, KingArticle, LamInPro2,LamInpro,PintoArticle}.
There are several challenges, however, for applying these techniques to CPSs. 
First, they operate on specific types of systems, mainly software 
systems, so more work is required to determine whether they are 
applicable to CPSs and whether they can recognize flaky test cases in testing CPSs. 
Second, these techniques focus on recognizing flaky test cases 
during regression testing, whereas flakiness can occur in CPSs 
at each level and in each type of testing. 
Third, these techniques usually rely on older versions of 
source code or test data from previous testing rounds, which 
provide pre-defined or learned anti-patterns in the test cases.
This, again, is possible in a regression testing setting
but not directly applicable to the scenarios we consider here.
Furthermore, this approach limits the techniques'
ability to a source-code-dependent subset of all possible flakiness causes. 
Finally, the cited work just investigates the flakiness of 
failed tests, whereas it is possible for a test of a CPS to
pass in a given execution and yet actually be flaky, a
possibility that our technique detects.

There has been work investigating the root causes of 
flaky behaviors in CPSs \cite{ZampettiInpro,ZampettiArticle} 
or embedded systems \cite{StrandbergInPro}, that provides
some solutions for maintaining CPSs after flaky tests are revealed, 
but the papers cited do not discuss approaches for recognizing flakiness. 

To the best of our knowledge, there has been no work 
on recognizing flaky tests for CPSs using any
approach other than ``Rerun''.

\section{Conclusion}
\label{sec:conclusion}

\noindent We have presented {\sc HyTestTF}, an approach for 
testing CPSs, that integrates the HyTest test case generaton and test oracle approach \cite{SadriHyTest} with additional steps that perform 
test results validation ({\sc TReVa}) and flaky test recognition ({\sc FlaRe}).
To evaluate the effectiveness of {\sc HyTestTF} in recognizing real failures, 
real passes, real flaky test cases, and real flaky modes, we 
studied its application on a set of CPSs and compared the results with 
the results obtained by applying {\sc HyTest} to all of these CPSs. 

Our results show that {\sc HyTestTF} was able to correctly validate 
the test results and differentiate test runs that failed under 
some condition(s) and passed under other condition(s) from those 
that passed or failed unconditionally.

We have studied {\sc HyTestTF} in relation to the testing of
CPSs at the MiL level; however, we believe that {\sc HyTestTF} 
could be applied to CPSs at the SiL and HiL levels as well to 
validate the test results and recognize flaky modes, 
flaky test conditions, and flaky test cases in those cases. 
Additional studies could be performed to assess this.

In our empirical study, we evaluated {\sc HyTestTF} on five 
CPSs of various (low, medium, high) complexity and our 
results are limited to these. 
Although we do not have any theoretical or practical 
reason to think that {\sc HyTestTF}'s applicability and 
scalability might be limited on larger and more complex
industrial CPSs, additional empirical work is needed to 
accurately determine this.

\begingroup
\raggedright
\bibliographystyle{IEEEtran}
\bibliography{main}{}
\endgroup

\end{document}